\documentstyle{article}
\font\tenbf=cmbx10
\font\tenrm=cmr10
\font\tenit=cmti10
\font\elevenbf=cmbx10 scaled\magstep 1
\font\elevenrm=cmr10 scaled\magstep 1
\font\elevenit=cmti10 scaled\magstep 1

\font\ninerm=cmr9

\font\eightrm=cmr8

\newcommand{\be}{\begin{equation}}
\newcommand{\ee}{\end{equation}}
\newcommand{\bea}{\begin{eqnarray}}
\newcommand{\eea}{\end{eqnarray}}
\newcommand{\del}{\partial}
\newcommand{\half}{{\textstyle \frac{1}{2}}}
\newcommand{\Int}[2]{\int\!\! d^{#1}\!{#2}}
\newcommand{\invint}[1]{\int\!\!\frac{d^3\! {#1}}{(2\pi)^3 2{#1}_0}}
\newcommand{\loopint}[1]{\int\!\!\frac{d^4\! {#1}}{(2\pi)^4}}
\newcommand{\k}{\mbox{\bf k}}
\newcommand{\fk}{{\mbox{\footnotesize\bf k}}}
\newcommand{\sP}[1]{{\mbox{\scriptsize\bf P}_{#1}}}
\newcommand{\sx}{{\mbox{\scriptsize\bf x}}}
\newcommand{\sk}{{\mbox{\scriptsize\bf k}}}
\newcommand{\adj}{{\scriptscriptstyle \dagger}}
\newcommand{\down}{a^{\rule{1mm}{0mm}}_{\fk \lambda}}
\newcommand{\up}{a^\adj_{\fk \lambda}}
\newcommand{\Down}{a^{\rule{1mm}{0mm}}_\lambda ({\bf k})}
\newcommand{\Up}{a^\adj_\lambda ({\bf k})}
\newcommand{\downp}{a^{\rule{1mm}{0mm}}_{\fk^\prime \lambda^\prime}}
\newcommand{\upp}{a^\adj_{\fk^\prime \lambda^\prime}}
\newcommand{\Downp}{a^{\rule{1mm}{0mm}}_{\lambda^\prime}({\bf k}^\prime)}
\newcommand{\Upp}{a^\adj_{\lambda^\prime}({\bf k}^\prime)}
\newcommand{\pol}{\varepsilon^\mu_{\fk \lambda}}
\newcommand{\polp}{\varepsilon^{\mu *}_{\fk \lambda}}
\newcommand{\polmu}[1]{\varepsilon^\mu_{\fk {#1}}}
\newcommand{\fkl}{f^{\rule{1mm}{0mm}}_{\fk \lambda}}
\newcommand{\ekl}{\eta^{\rule{1mm}{0mm}}_{\fk \lambda}}
\newcommand{\ham}[1]{H_{\mbox{\ninerm #1}}}
\newcommand{\ket}[1]{| #1 \rangle}
\newcommand{\bra}[1]{\langle #1 |}
\newcommand{\braket}[2]{\langle #1 | #2 \rangle}
\newcommand{\vac}{|\mbox{\rm vac}\rangle}
\newcommand{\Vac}{\langle\mbox{\rm vac}|}
\newcommand{\Evac}{|\mbox{\small\sc vac}\rangle}
\newcommand{\EVac}{\langle\mbox{\small\sc vac}|}
\newcommand{\x}{{\mbox{\rm x}}}
\newcommand{\y}{{\mbox{\rm y}}}
\newcommand{\q}{{\mbox{\rm q}}}
\newcommand{\D}{D_{\mbox{\scriptsize\eightrm x}}}
\newcommand{\bold}{\elevenbf}
\newcommand{\ital}{\elevenit}
\makeatletter
\@addtoreset{equation}{section}
\makeatother

\newcommand{\sect}[2]{\section*{\bold {#1}. {#2}}\setcounter{section}{#1}
 \setcounter{equation}{0}\vspace*{-5mm}\indent}
\renewenvironment{thebibliography}[1]
 { \elevenrm
   \begin{list}{\arabic{enumi}.}
    {\usecounter{enumi} \setlength{\parsep}{0pt}
     \setlength{\itemsep}{3pt} \settowidth{\labelwidth}{#1.}
     \sloppy
    }}{\end{list}}

\begin{document}
\begin{flushright}
ADP-95-30/T184 \\
hep-th/9505152
\end{flushright}

\begin{center}
{\large{\bf INTRODUCTION TO QUANTUM FIELD THEORY }} \\
\vspace{2.2 cm}
 R. J. CREWTHER \\
\vspace{1.2 cm}
{\it
Department of Physics and Mathematical Physics \\
University of Adelaide, S.Aust 5005, Australia } \\
\vspace{2.4 cm}
\begin{abstract}
Even the uninitiated will know that Quantum Field Theory cannot be
introduced systematically in just four lectures. I try to give a reasonably
connected outline of part of it, from second quantization to the path-integral
technique in Euclidean space, where there is an immediate
connection with the rules for Feynman diagrams and the partition function
of Statistical Mechanics.
\end{abstract}

\end{center}
\vspace{2.5cm}
To appear in the proceedings of the ``Seventh Physics Summer School--
Statistical Mechanics and Field Theory" (to be published by World
Scientific) held at the Australian National University.
\vspace{2.5cm}
\begin{flushleft}
E-mail: {\it rcrewthe@physics.adelaide.edu.au}
\end{flushleft}

\newpage

\parindent=1.5pc
\baselineskip=10pt
\begin{center}{{\tenbf INTRODUCTION TO QUANTUM FIELD THEORY\\}
\vglue 1.0cm
{\tenrm R. J. CREWTHER\\}
\baselineskip=13pt
{\tenit Department of Physics and Mathematical Physics,
University of Adelaide\\}
\baselineskip=12pt
{\tenit Adelaide, S.A. 5005, Australia\\}
\vglue 0.8cm
{\tenrm ABSTRACT}}
\end{center}
\vglue 0.3cm
{\rightskip=3pc
\leftskip=3pc
\tenrm\baselineskip=12pt
\noindent
Even the uninitiated will know that Quantum Field Theory cannot be
introduced systematically in just four lectures. I try to give a reasonably
connected outline of part of it, from second quantization to the path-integral
technique in Euclidean space, where there is an immediate
connection with the rules for Feynman diagrams and the partition function
of Statistical Mechanics.
\vglue 0.6 cm}
\baselineskip=14pt
\elevenrm
\sect{1}{Why Introduce Quantum Fields?}

In ordinary quantum mechanics, displacement is an operator $\bf X$,
but time $t$ is just a parameter which labels Schr\"{o}dinger state
vectors or Heisenberg operators such as ${\bf X = X}(t)$. This does
not sit well with special relativity, which places displacement and
time on the same footing.

If time were an operator $T\!$, it would be the component of a four-position
operator $X^\mu = (T,{\bf X})$ conjugate to the Hamiltonian $H$ in the
four-momentum\footnote{\ninerm Generally, we use natural
units $\hbar = c = \epsilon_0 = 1$. The metric tensor $g^{\mu\nu}$ is
$g^{00}=1,\ g^{0i}~=~0,\ g^{ij} = - \delta^{ij}$
for spatial indices $i,j = 1,2,3$. Thus the four-derivative
$\del^\mu = \del/\del x_\mu$ has components
$(\del/\del t, - {\bf \nabla})$. The sign of the antisymmetric tensor
$\varepsilon_{\alpha\beta\gamma\delta}$ is fixed by $\varepsilon_{0123}
= +1 = - \varepsilon^{0123}$.} $P^\mu = (H,{\bf P})$:
\be [P^\mu ,X^\nu] = ig^{\mu\nu} \ee
The commutator $[H,T] = i$ implies
\be \exp (-i\epsilon T)\, H \exp (i\epsilon T) = H - \epsilon \ee
for any constant $\epsilon$\/, so the operator
$\exp (i\epsilon T)$ applied to any eigenstate
$|E\rangle$ of $H$ with energy eigenvalue $E$ produces
another eigenstate
\[ \exp (i\epsilon T)\, |E\rangle  \]
with shifted eigenvalue $E-\epsilon$. That indicates the presence of
a continuous energy spectrum with range $-\infty < E < \infty$,
contrary to the requirement that $E$ be bounded below. Also,
it contradicts the fact that generally, $E$ is quantized \cite{Pauli}.

So instead, we demote displacement to the status of parameter,
like $t$. The dynamical quantities describing matter are to be
operators labelled by a spatial three-vector $\bf x$ as well as the time $t$.
Such operators are {\elevenit quantum fields}
\be  \phi = \phi (x) \ee
where $x$ denotes the four-vector parameter $(t,{\bf x})$.
\sect{2}{Photons}

Experimental evidence for field quantization dates back to the
discovery of photons. The success of Einstein's analyses of the
photoelectric effect in 1905 and photon emission and absorption
in 1917 \cite{Einstein} contradicted purely classical interpretations
of the electromagnetic four-potential $A_\mu(x)$ and field strength
tensor
\be F_{\mu\nu} = \del_\mu A_\nu - \del_\nu A_\mu  \ee

With the birth of quantum mechanics in 1925--26, it became evident
that $F_{\mu\nu}$ and $A_{\mu}$ must be operators acting on state
vectors, like other dynamical variables, with Maxwell's equations
understood to be operator relations:
\be \del^\mu F_{\mu\nu} = j_\nu \hspace{3mm} , \hspace{3mm}
    \del^\mu \widetilde{F}_{\mu\nu} = 0    \label{2a}   \ee
Here $\widetilde{F}_{\mu\nu} = \frac{1}{2}\varepsilon_{\mu\nu\alpha\beta}
F^{\alpha\beta}$ is the dual tensor of $F_{\alpha\beta}$, and $j_\mu(x)$
is the conserved four-current density due to charged matter ($\del^\mu
j_\mu = 0$).

Quantum electrodynamics began in 1927 with Dirac's paper \cite{Dirac} on the
operator structure of the Hamiltonian. His method has become known as
``second quantization''.

Consider photons within a large box of volume $V = L_1 \times L_2 \times L_3$.
Periodic boundary conditions in each spatial dimension restrict the
allowed wave vectors to values
\be  \k = 2\pi\left(m_1/L_1,m_2/L_2,m_3/L_3\right)  \label{2a0}  \ee
where $m_1,m_2,m_3$ are integers. Continuously varying $\k$ can be
obtained in the limit $V\rightarrow\infty$.

According to Einstein, these photons have electromagnetic energy
\[ \sum_\fk \sum^2_{\lambda=1} |\k|\,n_{\fk \lambda}  \]
where $n_{\fk \lambda}$ is the number of photons with wave vector
$\k$ and transverse polarization $\lambda$, and $\sum_\fk$ means
sum over $m_1,m_2,m_3$. Dirac understood the integers
$n_{\fk \lambda} \geq 0$ to be eigenvalues of a quantum number
operator $N_{\fk \lambda}$, and so obtained a Hamiltonian operator for
photons:
\be
\ham{photon} = \sum_\fk \sum^2_{\lambda=1} |\k|\,N_{\fk \lambda}
\ee

How should the quantization of $N_{\fk \lambda}$ be achieved? The
operator $-i\del/\del\theta$ conjugate to an angle variable
$\theta$ with period $2\pi$ has integer eigenvalues, but these run from
$+\infty$ down to $-\infty$, contrary to the requirement that energy be
bounded below.

Dirac observed that the required eigenvalue spectrum can be obtained from
the algebra of raising and lowering operators for independent harmonic
oscillators. Consider operators $\down$ and $\up$ with oscillator-like
commutation relations
\be
[\down , \upp ] = \delta_{\fk\fk^\prime}\delta_{\lambda\lambda^\prime}
\hspace{5mm},\hspace{5mm} [\down , \downp ] = 0 = [\up , \upp ]
\label{2a1}  \ee
Then it is possible to make the identification
\be  N^{\rule{0mm}{1mm}}_{\fk\lambda} = \up \down  \ee
because we get integer eigenvalues running from 0 to $\infty$. The result
for the photon Hamiltonian is a quadratic form
\be \ham{photon} = \sum_\fk \sum^2_{\lambda=1} |\k|\,\up\down \label{2b} \ee
with minimum eigenvalue zero corresponding to a state $|0\rangle$
containing no photons. Since $\down$ reduces the number of
photons of type $(\k,\lambda)$ by one, it is called a photon
annihilation operator. Similarly, $\up$ ``creates'' a photon.

The theory (\ref{2b}) describes free photons only, since it implies
that all number operators are conserved:
\be  [N^{\rule{0mm}{1mm}}_{\fk\lambda},\ham{photon}] = 0  \ee
Dirac included photon interactions by adding a linear combination of
$\down$ and $\up$ to correspond to the absorption or emission of a
single photon in first-order perturbation theory:
\be
\ham{int} = \sum_{\fk}\sum^2_{\lambda=1}\left(V^\adj_{\fk\lambda}\down
              + \up V^{\rule{0mm}{1mm}}_{\fk\lambda}\right)
\label{2c}  \ee
The factors $V^{\rule{0mm}{1mm}}_{\fk\lambda}$ and $V^\adj_{\fk\lambda}$
depend on the dynamical variables characterizing charged matter, such
as atoms. The full Hamiltonian is then of the form
\be  H = \ham{photon} + \ham{atom} + \ham{int}  \ee
where $\ham{atom}$ describes the unperturbed atomic system.

This theory is not fully explicit --- it specifies the dependence of $H$ on
$\down$ and $\up$ but not on the dynamical variables
for charged matter. Nevertheless Dirac was able to derive from his
theory the key points of Einstein's analysis of photon emission and
absorption, and hence the $A$ and $B$ coefficients. Briefly, the relations
\be \up |\ldots n_{\fk\lambda}\ldots\rangle
   = \sqrt{n_{\fk\lambda}+1}\,|\ldots n_{\fk\lambda+1}\ldots\rangle
\hspace{5mm},\hspace{5mm}
 \down |\ldots n_{\fk\lambda}\ldots\rangle
   = \sqrt{n_{\fk\lambda}}\,|\ldots n_{\fk\lambda}\ldots\rangle  \ee
imply that, given $n_{\fk\lambda}$ photons initially, the probabilities of
emission and absorption are given by
\be {\cal P}_{\mbox{\ninerm em}}
  = \left(n_{\fk\lambda}+1\right){\cal P}_{\fk\lambda} \ee
and
\be {\cal P}_{\mbox{\ninerm abs}} = n_{\fk\lambda}{\cal P}_{\fk\lambda}
\ee
where ${\cal P}_{\fk\lambda}$ is the emission probability in the absence
of photons. Thus the emission probability consists of a spontaneous part
${\cal P}_{\fk\lambda}$, plus an induced part
$n_{\fk\lambda}{\cal P}_{\fk\lambda}$ equal to the absorption
probability, in agreement with Einstein \cite{Einstein}.
\newpage\noindent
\sect{3}{Electromagnetic Field Operators}

The photon operators $\down$ and $\up$ can be associated with the
corresponding Fourier components and polarizations of $F^{\mu\nu}$
and $A^\mu$ if these fields are {\ital free\/}. This means that
charged matter is omitted in Maxwell's equations (\ref{2a}), at
least in an initial approximation:
\be \del^\mu F_{\mu\nu} = 0 \hspace{3mm} , \hspace{3mm}
    \del^\mu \widetilde{F}_{\mu\nu} = 0    \label{3a}   \ee

At this point, we have to deal with the partly unphysical nature of the
four-potential $A_\mu$. Photons are polarized transversely, i.e. in only
two directions $\lambda~=~1,2$. The remaining ``longitudinal'' and
``scalar'' components of $A_\mu$ must be sidelined in some manner. This
is done by exploiting gauge ambiguities in the four-potential:
\be A_\mu \longrightarrow A_\mu + \del_\mu\xi \hspace{5mm}
           \mbox{\rm for any}\ \xi=\xi(x)  \label{3b}  \ee
These transformations are an exact invariance of electromagnetism, so
they act within a completely unphysical sector of the theory.\\[4mm]
{\ital 3.1 Unphysical \vspace{1mm}Components}

There are various ways of proceeding, depending on
taste\footnote{See Chapter 14 of Bjorken and Drell \cite{BjD2}.
Lee \cite{Lee} and Ryder \cite{Ryder} quantize directly in the Coulomb
gauge. Bogoliubov and Shirkov \cite{BS}, Schweber \cite{SSS},
Itzykson and Zuber \cite{IZ}, and Mandl and Shaw \cite{MS} use the
Lorentz covariant procedure of Gupta and Bleuler \cite{GB}, where
Hilbert space is expanded to include unphysical states of zero and
negative norm.}.
One is to use (\ref{3a}) and (\ref{3b}) to eliminate the two extra
components of $A_\mu$ completely.

For each potential $A_\mu$, consider the $A_\mu$-dependent gauge
transformation
\be
A_\mu \longrightarrow {\cal A}_\mu(t,{\bf x}) = A_\mu(t,{\bf x})
     - \frac{\del}{\del x^\mu}\Int{3}{y}\:\frac{1}{4\pi|{\bf x-y}|}
       \nabla_{\bf y}\cdot{\bf A}(t,{\bf y})
\ee
A shorthand for this is
\be {\cal A}_\mu = A_\mu + \del_\mu \nabla^{-2}\nabla\cdot{\bf A} \ee
where
\be \nabla^{-2}f(t,{\bf x})
= - \frac{1}{4\pi}\Int{3}{y}\:\frac{f(t,{\bf y})}{|{\bf x-y}|}
\label{3b0}\ee
defines the action of the inverse Laplacian $\nabla^{-2}$ on a function
$f$. Convergence of (\ref{3b0}) imposes a boundary condition on
potentials at space-like infinity such that zero modes of $\nabla^{-2}$
\be  a_\mu(x) = x^\nu f_{\mu\nu}(x_0)  \ee
are excluded. Then both $\nabla^2\nabla^{-2}=1$ and $\nabla^{-2}\nabla^2=1$
are legitimate operator identities.

The new three-vector potential
\be  {\cal A} = {\bf A} - \nabla\nabla^{-2}\nabla\cdot{\bf A}  \ee
obeys the Coulomb gauge condition
\be  \nabla\cdot{\cal A} = 0  \ee
so it is transverse: its Fourier components are orthogonal to the wave vector
${\bf k}$. The new scalar potential is
\be  {\cal A}_0 = A_0 + \del_0 \nabla^{-2}\nabla\cdot{\bf A}  \ee
Eq.\ (\ref{3a}) contains Gauss's Law for free fields
\be  \del^\mu F_{\mu 0} = - \nabla^2 A_0 - \del_0\nabla\cdot{\bf A} = 0 \ee
which requires the temporal gauge condition
\be  {\cal A}_0 = 0   \label{3b1}  \ee
to be obeyed as well.

The lack of Lorentz covariance of this procedure is only apparent, since
$A_\mu$ is not a true four-vector --- any Lorentz transformation
$x \rightarrow x^\prime = \Lambda x$ may be accompanied by a gauge
transformation $\xi_\Lambda$ with arbitrary dependence on $\Lambda$:
\be  A_\mu(x) \longrightarrow A^\prime_\mu(x^\prime)
   =  \Lambda^{\ \nu}_\mu A_\nu(x) + \del_\mu \xi_\Lambda (x)  \ee

More generally, let $N_\mu$ be any constant time-like vector with
$N^2=1$, and let
\be  \left( (N\cdot\del)^2 - \del^2\right)^{-1}f(x) =
  - \Int{4}{y}\:\delta(N\cdot x - N\cdot y)f(y)/\sqrt{- 16\pi^2(x-y)^2} \ee
Then free fields
\be  {\cal A}_\mu = A_\mu + \del_\mu\left((N\cdot\del)^2 - \del^2\right)^{-1}
                    (\del\cdot A - N\cdot\del N\cdot A)  \ee
satisfy transversality conditions
\be  \del\cdot{\cal A} = 0 \ ,\ \ N\cdot{\cal A} = 0  \label{3c}  \ee
The choice of $N_\mu$ has no physical significance: any two $N$'s can be
related by a gauge transformation.\\[4mm]
{\ital 3.2 Fourier \vspace{1mm}Decomposition}

The field strengths $F_{\mu\nu}$ are gauge invariant, so we have
\be
F_{\mu\nu} = \del_\mu {\cal A}_\nu - \del_\nu {\cal A}_\mu
\label{3d}  \ee
Eqs.\ (\ref{3c}) and (\ref{3d}) imply the equation of motion
\be  \del^2 {\cal A}_\mu = 0  \label{3d1} \ee
and hence the Fourier decomposition
\be  {\cal A}^\mu  =  \sum_\fk\sum^2_{\lambda = 1}
\left\{ c^{\rule{1mm}{0mm}}_{\fk \lambda} \pol e^{-ik\cdot x}
     + c^\adj_{\fk \lambda} \polp e^{ik\cdot x} \right\}  \label{3e}  \ee
in which the variable $k^\mu$ conjugate to $x_\mu$ lies on the forward
light cone
\be  k^\mu = (|{\bf k}|, {\bf k})  \label{3e1}  \ee
and the four-vectors $\pol$ are orthogonal to both $k_\mu$ and $N_\mu$:
\be  k\cdot\varepsilon^{\rule{1mm}{0mm}}_{\fk \lambda}
  = 0 = N\cdot\varepsilon^{\rule{1mm}{0mm}}_{\fk \lambda}  \ee
For the moment, the coefficients $c^{\rule{1mm}{0mm}}_{\fk \lambda}$ and
$c^\adj_{\fk \lambda}$ remain unspecified, except that they are operators
chosen to make ${\cal A}_\mu$ Hermitian.

The polarization vectors $\pol$ are space-like, and can be chosen
orthogonal to each other:
\be \varepsilon^{\rule{1mm}{0mm}}_{\fk \lambda}\cdot
 \varepsilon^{*}_{\fk \lambda^\prime} = - \delta_{\lambda\lambda^\prime}\ ,
 \ \ \lambda,\lambda^\prime = 1,2   \label{3e0}  \ee
For example, if we take
\be  N_\mu = (1,0,0,0)\ , \ \ k_\mu = (k,0,0,k)  \ee
we can choose $\pol$ to be
\be \polmu{x} = (0,1,0,0)\ , \ \ \polmu{y} = (0,0,1,0) \label{3e2} \ee
for plane polarizations, or
\be \polmu{\pm} = \frac{1}{\sqrt{2}}(0,1,\pm i,0)  \label{3e3} \ee
for circular polarizations. The values $\lambda = 1,2$ refer to any
orthogonal pair of polarizations, such as (\ref{3e2}) or (\ref{3e3}).

The set
\be \left\{ N^\mu\: ,\: N^\mu - k^\mu/k\cdot N\: , \: \polmu{1}\: ,
                                      \: \polmu{2} \right\}
%\label{3e4}
\ee
is orthonormal and complete in four-dimensional space with metric
$g_{\mu\nu}$, so we have
\be  g^{\mu\nu} = N^\mu N^\nu
    - (N^\mu - k^\mu/k\cdot N)(N^\nu - k^\nu/k\cdot N)
    - \sum^2_{\lambda = 1} \pol \varepsilon^{\nu *}_{\fk \lambda}  \ee
and hence
\be  \sum^2_{\lambda = 1} \pol \varepsilon^{\nu *}_{\fk \lambda}
= - g^{\mu\nu} + (N^\mu k^\nu + N^\nu k^\mu)/k\cdot N
   - k^\nu k^\nu/(k\cdot N)^2   \label{3e5} \ee
The presence of the unphysical entity $N^\mu$ in this formula is related to
the lack of gauge invariance of the polarization vectors
\be  \pol \longrightarrow \pol + k^\mu \xi_\lambda(k)\ ,
                                       \ \ \lambda=1,2  \label{3f} \ee
This occurs even though the label $\lambda$ refers to transverse directions.
Any calculation of a physical amplitude should be invariant under
(\ref{3f}) and hence independent of artefacts like $N_\mu$.\\[4mm]
{\ital 3.3 Connection with \vspace{1mm}Photons}

The summation variable ${\bf k}$ in the Fourier expansion (\ref{3e}) of
${\cal A}^\mu$ is an obvious candidate for photon momentum, but the
precise identification is not immediately clear. Does a photon of
four-momentum $k_\mu$ correspond to a real wave
\be  \sin(k\cdot x + {\rm phase})  \label{3f1}\ee
as suggested by classical radiation theory, or to a complex de Broglie
wave $e^{-ik\cdot x}$, and if the latter, what does $e^{+ik\cdot x}$
represent?

To resolve this question, consider a photon being absorbed by a heavy
neutral system with charged constituents, such as an atom or a brick:
\be \{\mbox{\rm atomic state $i$}\} + \gamma({\bf k},\lambda)
      \longrightarrow \{\rm \mbox{atomic state $f$}\}  \label{3f2}\ee
This process is induced by the Hamiltonian operator

\be  \ham{int}(t) = \Int{3}{x}\: j^\mu(x){\cal A}_\mu(x)  \label{3g}  \ee
We consider first-order perturbations, so it is legitimate to use the
free field ${\cal A}_\mu$ in (\ref{3g}). Also, we choose the temporal
gauge (\ref{3b1}). According to time-dependent perturbation theory
\cite{Dirac,Baym,Sakurai}, the amplitude for the transition (\ref{3f2}) is
\be {\cal S}_{i + \gamma \rightarrow f}
= -i\:\bra{f}\int\!\! dt\, \ham{int}(t)\,\ket{i,\gamma}
= -i\!\Int{4}{x}\,\bra{f}\vec{j}(x)\ket{i}\cdot
          \bra{0}\vec{\cal A}(x)\ket{\gamma}
\label{3g1} \ee

If the system is sufficiently massive, both $\ket{i}$ and $\ket{f}$ can
be represented by non-relativistic wave functions
\be  \psi_i(t,{\bf x}) = e^{i(\sP{i}\cdot\sx - E_i t)} \chi_i  \ee
and
\be  \psi_f(t,{\bf x}) = e^{i(\sP{f}\cdot\sx - E_f t)} \chi_f  \ee
where $E_f,E_i$ and ${\bf P}_f,{\bf P}_i$ denote centre-of-mass
energies and momenta. The spinors $\chi_i$ and $\chi_f$ depend on
the internal dynamical variables to which $\vec{j}$ couples. All
dependence on the atomic centre-of-mass coordinate $x^\mu = (t,{\bf x})$
is carried by $\psi_i$ and $\psi^\adj_f$:
\be \bra{f}\vec{j}(x)\ket{i} =
      e^{i(E_f - E_i)t} e^{-i(\sP{f} - \sP{i})\cdot\sx}
             \bra{f}\vec{j}(0)\ket{i}
\ee

Consequently, the amplitude (\ref{3g1}) can be written
\be {\cal S}_{i + \gamma \rightarrow f}
= -i\Int{4}{x}\, e^{i(E_f - E_i)t} e^{-i(\sP{f} - \sP{i})\cdot\sx}\:
\bra{0}\vec{\cal A}(x)\ket{\gamma}\cdot\bra{f}\vec{j}(0)\ket{i}
%\label{3h}
\ee
Clearly, only one Fourier component in the formula (\ref{3e}) for
${\cal A}^\mu$ contributes to ${\cal S}_{i + \gamma \rightarrow f}$:
the term proportional to $e^{-ik\cdot x}$ with
\be  k^0 = |{\bf k}| = E_f - E_i\ , \ \ {\bf k} = {\bf P}_f - {\bf P}_i
\label{3h1} \ee
Therefore the amplitude to
absorb a photon of type $({\bf k}, \lambda)$ is
\be {\cal S}_{i + \gamma \rightarrow f}
= -(2\pi)^4 i\delta^4(P_f - P_i - k)\, f^{\rule{1mm}{0mm}}_{\fk \lambda}
 \vec{\varepsilon}^{\rule{1mm}{0mm}}_{\fk\lambda}\!
                             \cdot\bra{f}\vec{j}(0)\ket{i}
\ee
where the amplitudes $f^{\rule{1mm}{0mm}}_{\fk \lambda}$ are given by
\be  f^{\rule{1mm}{0mm}}_{\fk \lambda}
   = \bra{0} c^{\rule{1mm}{0mm}}_{\fk \lambda} \ket{\gamma({\bf k},\lambda)}
\ee
By absorbing phases into the definition of $\ket{\gamma({\bf k},\lambda})$,
we can require the {\elevenit c}-numbers $f^{\rule{1mm}{0mm}}_{\fk \lambda}$
to be real and positive.

Evidently the coefficient $c^{\rule{1mm}{0mm}}_{\fk \lambda}$ in the
Fourier series (\ref{3e}) is an operator which reduces photon number
by one. For if the initial state had contained $n$ photons, only that
photon obeying the constraint (\ref{3h1}) could have been absorbed in
first-order perturbation theory, leaving $n-1$ photons to continue on
to the final state. Therefore $c^{\rule{1mm}{0mm}}_{\fk \lambda}$ must
be proportional to the photon annihilation operator $\down$:
\be  c^{\rule{1mm}{0mm}}_{\fk \lambda} = \fkl\down  \label{3i} \ee

The interacting photon transmits both orbital and spin angular momentum
to the target atom. Generally, this is analyzed by decomposing
$\vec{\cal A}(x)$ into multipole fields \cite{Baym,Brink,RoseEdmonds}. The
simplest case is $S$-wave absorption, which dominates for inverse photon
energy $|{\bf k}|^{-1}$ much larger than the dimensions of the target atom.
A further simplification is to consider atomic transitions
$\ket{j,m_i}\rightarrow\ket{j,m_f}$ in which the total angular
momentum quantum number $j$ of the atom is unchanged. Then the
Wigner-Eckart theorem reduces to the projection theorem for vector
operators \cite{Baym,Sakurai,Brink}
\be \bra{f}\vec{j}\ket{i} \propto \bra{f}\vec{J}\,\ket{i} \ee
where $\vec{J}$ is the operator for total atomic angular momentum:
\be {\cal S}_{i + \gamma \rightarrow f}
\propto \Int{4}{x}\, e^{i(E_f - E_i)t} e^{-i(\sP{f} - \sP{i})\cdot\sx}\:
\bra{0}\vec{\cal A}(x)\ket{\gamma}\cdot\bra{f}\vec{J}\,\ket{i}
%\label{3i1}
\ee
In the frame (\ref{3e1}), circular polarizations (\ref{3e3}) give rise
to raising and lowering operators for angular momentum:
\be  \vec{\varepsilon}^{\rule{1mm}{0mm}}_{\fk\pm}\cdot\vec{J}
  = \frac{1}{\sqrt{2}}(J_x \pm i J_y)  \ee
We see that the absorption of a photon with polarization $\polmu{+}$
increases the atomic spin component $J_z$ by 1, so the photon
is right-handed and carries spin (helicity) 1.

Photon emission can be analysed in the same way. The de Broglie wave
$\polp e^{ik\cdot x}$ in (\ref{3e}) corresponds to a photon of type
$({\bf k}, \lambda)$ being emitted. Substituting (\ref{3i}), we find
\be  {\cal A}^\mu
=  \sum_\fk\sum^2_{\lambda = 1} f^{\rule{1mm}{0mm}}_{\fk \lambda}
\left(\down \pol e^{-ik\cdot x} + \up \polp e^{ik\cdot x} \right)
\label{3j}  \ee
Note the consistency between this result and Eqs.\ (\ref{2c}) and
(\ref{3g}) for $\ham{int}$.\\[4mm]
{\ital 3.4 Electromagnetic Energy and \vspace{1mm}Momentum}

The constants $\fkl$ in (\ref{3j}) can be determined by comparing Dirac's
photon energy operator (\ref{2b}) with Maxwell's formula for
electromagnetic energy
\be
\ham{emag} = \half\Int{3}{x}
             \left({\bf E}^2 + {\bf B}^2\right)   \label{3k} \ee
where ${\bf E}$ and ${\bf B}$ are the electric and magnetic fields.

Eqs.\ (\ref{3d}) and (\ref{3j}) imply
\bea
E_i &=& F_{0i}\ =\ -i\sum_{\fk}\sum^{2}_{\lambda = 1}\fkl\down
         \left(k_0 \varepsilon_i - k_i \varepsilon_0\right)_{\fk\lambda}
         e^{-ik\cdot x}\ +\ \mbox{\rm h.c.} \nonumber \\
B_i &=& - \half\varepsilon_{ijk}F_{jk}
   \ =\ i\sum_{\fk}\sum^{2}_{\lambda = 1}\fkl\down\varepsilon_{i\ell m}
         \left(k_\ell \varepsilon_m\right)_{\fk\lambda}
         e^{-ik\cdot x}\ +\ \mbox{\rm h.c.}
\eea
where ``h.c.'' denotes Hermitian conjugate, so $\Int{3}{x}\,{\bf E}^2$
and $\Int{3}{x}\,{\bf B}^2$ are sums $\sum_{\fk\fk^\prime}
\sum_{\lambda\lambda^\prime}$ containing integrals
\be  \Int{3}{x}\, e^{i\sk\cdot\sx}e^{\mp i\sk^\prime \cdot\sx}
    =  V\,\delta_{\fk ,\pm \fk^\prime}  \ee
For ${\bf k}^\prime = - {\bf k}$ contributions, a convention is needed to
fix relative phases in  $\varepsilon^\mu_{\pm\fk\,\lambda}$ and hence
$a^{\rule{1mm}{0mm}}_{\pm\fk\,\lambda}$. In terms of circular
polarizations $\lambda = \pm$, a simple choice is
\be  \varepsilon^\mu_{-\fk\,\pm} = \varepsilon^{\mu *}_{\fk\,\pm}  \ee
Then Eqs.\ (\ref{3e1})--(\ref{3e0}) imply
\bea
\Int{3}{x}\:{\bf E}^2 &=& V\sum_{\fk}\sum_{\lambda=\pm}k^2_0\fkl\down
  \left(\fkl\up - f^{\rule{1mm}{0mm}}_{-\fk\lambda}
   a^{\rule{1mm}{0mm}}_{-\fk\lambda}e^{-2ik_0t}\right)\ +\ \mbox{\rm h.c.}
\nonumber \\
\Int{3}{x}\:{\bf B}^2 &=& V\sum_{\fk}\sum_{\lambda=\pm}k^2_0\fkl\down
  \left(\fkl\up + f^{\rule{1mm}{0mm}}_{-\fk\lambda}
   a^{\rule{1mm}{0mm}}_{-\fk\lambda}e^{-2ik_0t}\right)\ +\ \mbox{\rm h.c.}
\eea
The ${\bf k}^\prime = - {\bf k}$ terms depend on time and do not conserve
photon number, but, as might be expected, they cancel in the
combination (\ref{3k}):
\bea  \ham{emag} &=& V \sum_{\fk}\sum_{\lambda=\pm}\left(k_0\fkl\right)^2
          \left(\down\up + \up\down\right)  \nonumber \\
&=&  2V \sum_{\fk}\sum_{\lambda=\pm}\left(k_0\fkl\right)^2
          \left(\up\down + \half\right)  \label{3l} \eea
Here the labels $\lambda = 1,2$ refer to any pair of transverse
polarizations obeying Eq.\ (\ref{3e0}).

The constant $\frac{1}{2}$ added to $\up\down$ in (\ref{3l}) is a
consequence of the commutation relations (\ref{2a1}). It is therefore
not surprising that Dirac's quantum formula (\ref{2b}) and the classical
expression (\ref{3k}) should differ by such terms. Agreement in the
classical limit $n_{\fk\lambda} \gg 1$ is achieved by equating the
coefficients of $\up\down$:
\be  \fkl = \frac{1}{\sqrt{2k_0 V}}\ , \ \ k_0 = |{\bf k}|  \label{3m}  \ee
For quantum mechanical purposes, the constant term in $\ham{emag}$ must be
subtracted off so that the no-photon state $\ket{0}$ has zero energy. The
result is (of course) Dirac's Hamiltonian
\be  \ham{photon} = \sum_{\fk}\sum^2_{\lambda = 1}\, k_0\,\up\down
= \ham{emag} - \sum_{\fk}\sum^2_{\lambda = 1}\,{\textstyle \frac{1}{2}} k_0
\label{3m0} \ee
This formula is often written in {\ital normal ordered\/} form
\be   \ham{photon} = \half\Int{3}{x}\: : {\bf E}^2 + {\bf B}^2 :
\label{3m1} \ee
where the symbol \, :\,\ldots\,:\, is an instruction to order all products
such that creation operators appear to the left of all annihilation
operators.

The subtraction
\be  \sum_{\fk}\sum^2_{\lambda = 1}\, \half k_0  \ee
is an infinite zero-point energy produced by the oscillator-like
formalism (\ref{2a1}). Such infinite ambiguities arise frequently in
quantum field theory. They are a consequence of attempts to multiply
field operators such as ${\bf E}(x)$ and ${\bf B}(x)$ evaluated at the
{\ital same\/} space-time point $x$. Normal ordering is a special case
\cite{Wightman} of a general procedure known as {\ital renormalization\/}
in which each ambiguity is eliminated by
imposing a common-sense physical requirement --- in the case above,
that $\ket{0}$ should have zero energy.

Eq.\ (\ref{3m}) completes the specification of the operator structure
of the free field (\ref{3j}):
\be  {\cal A}^\mu = \sum_\fk\sum^2_{\lambda = 1} \frac{1}{\sqrt{2k_0 V}}
\left\{\down \pol e^{-ik\cdot x} + \up \polp e^{ik\cdot x} \right\}
\label{3n}  \ee

We can check Eq.\ (\ref{3n}) by calculating the Poynting vector
${\bf E \times B}$ and relating it to the three-momentum operator
${\bf P}$:
\be  {\bf P} = \Int{3}{x}\: :{\bf E \times B} :\
      = \sum_\fk\sum^2_{\lambda = 1} {\bf k}\,\up\down  \label{3o}  \ee
Here the ${\bf k}^\prime = -{\bf k}$ terms sum to zero because
\be  -\half\sum_\fk\sum^2_{\lambda = 1}
\down a^{\rule{1mm}{0mm}}_{-\fk\lambda}\,{\bf k}\, e^{-2ik_0 t}\
+\ \mbox{\rm h.c.}   \ee
changes sign for ${\bf k} \rightarrow -{\bf k}$. The same argument is not
really adequate for the ill-defined commutator term
\be \sum_{\fk}\sum^2_{\lambda = 1}\, \half {\bf k}  \ee
so normal ordering is required in Eq.\ (\ref{3o}).

The total four-momentum operator
\be  P^\mu = \sum_\fk\sum^2_{\lambda = 1} k^\mu \,\up\down  \label{3o1} \ee
satisfies the following commutation relations,
\be  \left[ P^\mu , \down\right] = - k^\mu \down \hspace{5mm},\hspace{5mm}
     \left[ P^\mu , \up\right] = k^\mu \up   \ee
so its commutator with ${\cal A}^\nu(x)$ corresponds to an infinitesimal
displacement in space-time:
\be  i \left[ P^\mu , {\cal A}^\nu \right] = \del^\mu {\cal A}^\nu  \ee
Finite displacements $x_\mu \rightarrow x_\mu + a_\mu$ can be obtained
via a unitary transformation:
\be e^{iP\cdot a} {\cal A}^\mu(x) e^{-iP\cdot a} =  {\cal A}^\mu(x+a)
\vspace{4mm} \label{3p}  \ee
{\ital 3.5 Classical \vspace{1mm}Waves}

The free field ${\cal A}_\mu(x)$ given by Eq.\ (\ref{3n}) is obviously not
diagonal in photon number. For example, it connects one-photon states to
$\ket{0}$, but there is no expectation value for a single photon:
\be  \bra{\gamma}{\cal A}_\mu\ket{\gamma} = 0  \ee
Indeed, we have seen that single photons are associated with de Broglie
waves ($e^{ik\cdot x}$ for emission, $e^{-ik\cdot x}$ for absorption)
and not with the real waves (\ref{3f1}) of classical electromagnetism.

Real waves can be obtained as field expectation values formed with
{\ital coherent states\/} \cite{IZ,coherent}
\be  \ket{\ekl} = \exp\left(-\half|\ekl|^2\right)\,\exp(\ekl\up)\,\ket{0}  \ee
which are eigenstates of $\down$:
\be  \down\ket{\ekl} = \ekl\ket{\ekl}  \label{3q}  \ee
Since $\down$ is not self-adjoint, its eigenvalues $\ekl$ can be complex
and its eigenstates non-orthogonal:
\be  \langle\ekl |\eta^\prime_{\fk\lambda}\rangle
= \exp\left(i\,{\rm Im}\,(\eta^{*}_{\fk\lambda} \eta^\prime_{\fk\lambda})
    - \half |\ekl - \eta^\prime_{\fk\lambda}|^2\right)  \ee
However, states with different $({\bf k},\lambda)$ values are orthogonal:
\be  \langle\ekl|\eta^{\rule{1mm}{0mm}}_{\fk^\prime\lambda^\prime}\rangle
   =  \delta_{\fk\fk^\prime}\delta_{\lambda\lambda^\prime}  \ee

The expectation value of ${\cal A}_\mu$ can be obtained directly
from Eq.\ (\ref{3q}):
\be  \bra{\ekl}{\cal A}^\mu(x)\ket{\ekl} = \sqrt{\frac{2}{k_0 V}}\,
        {\rm Re}\left\{\ekl\pol e^{-ik\cdot x}\right\}  \ee
These real Fourier components may be superposed by considering
expectation values $\bra{\zeta ,\eta}{\cal A}^\mu\ket{\zeta ,\eta}$
formed with states
\be  \ket{\zeta ,\eta}
= \sum_{\fk\lambda}\zeta^{\rule{1mm}{0mm}}_{\fk\lambda}\ket{\ekl}\ , \ \
\sum_{\fk\lambda}\left|\zeta^{\rule{1mm}{0mm}}_{\fk\lambda}\right|^2 = 1
\vspace{4mm}  \ee
{\ital 3.6 \vspace{1mm}Remarks}

So far, we have treated the wave vector ${\bf k}$ as a discrete variable
(\ref{2a0}). As the quantization volume becomes infinite, sums and
Kronecker deltas are replaced by integrals and Dirac delta functions:
\bea  (2k_0 V)^{-1}\sum_{\fk} &\longrightarrow&
               \invint{k}\ \ , \ \ \ \ k_0 = |{\bf k}| \nonumber \\
  2k_0 V\,\delta_{\fk\fk^\prime} &\longrightarrow&
                   (2\pi)^3 2k_0 \delta^3({\bf k}-{\bf k}^\prime) \eea
The extra factors $2k_0$ are conventional; they make the continuum result
Lorentz invariant, as is evident from the identity
\be
d^3\! k/2\sqrt{{\bf k}^2 + m^2} = d^4\! k\, \delta(k^2 - m^2)\theta(k_0)
\label{3qa}  \ee
with $m=0$ for photons.

 From Eq.\ (\ref{2a1}), we see that the $V\rightarrow\infty$ limit for
creation and annihilation operators is given by
\be   (2k_0 V)^{-1/2}\down \longrightarrow \Down    \ee
where the new operators satisfy
\be  [\Down , \Downp ] = 0 = [\Up , \Upp ]\ \ , \ \ \ [\Down , \Upp ] =
(2\pi)^3 2k_0 \delta^3({\bf k}-{\bf k}^\prime)\delta_{\lambda\lambda^\prime}
\label{3r}  \ee
The right-hand side of Eq.\ (\ref{3r}) corresponds to the normalization
\be
\langle\gamma({\bf k},\lambda)|\gamma({\bf k}^\prime,\lambda^\prime)\rangle
=(2\pi)^3 2k_0 \delta^3({\bf k}-{\bf k}^\prime)\delta_{\lambda\lambda^\prime}
\ee
for one-photon states. Photon polarization vectors $\pol$ are unaffected by
the limiting procedure.

Most discrete-${\bf k}$ formulas may be readily converted by applying these
rules. For example, Eqs.\ (\ref{3n}) and (\ref{3o1}) for the free field
${\cal A}^\mu(x)$ and the four-momentum operator $P^\mu$ become
\be  {\cal A}^\mu(x) = \sum^2_{\lambda = 1}\invint{k} \left\{
      \Down\pol e^{-ik\cdot x} + \Up\polp e^{ik\cdot x}\right\} \label{3s} \ee
and
\be  P^\mu = \sum^2_{\lambda = 1}\invint{k} \Up\Down\, k^\mu \label{3t} \ee

It is important to realise that equations such as (\ref{3s}) and (\ref{3t})
are valid only for free fields. An interacting four-potential $A^\mu$
satisfies
\be  (\del^2 g_{\mu\nu} - \del_\mu\del_\nu)A^\mu = j_\nu  \ee
instead of Eq.\ (\ref{3d1}), so it has Fourier components which do
{\ital not\/} satisfy the constraint $k^2 = 0$. Consequently, its
Fourier coefficients are not simply creation and annihilation operators
for single photons --- they can create or destroy multiparticle states.

However, the translation property (\ref{3p}) is generally satisfied by
field operators in local quantum field theories.
\sect{4}{Field Commutators and Uncertainty}

The field-strength tensor $F_{\mu\nu}$ is an operator, so its components
need not and in general do not commute. Consequently, measurements of
electric and magnetic fields are governed by uncertainty relations which
(for the moment) we write in the naive form
\be \Delta F_{\mu\nu}(x)\, \Delta F_{\alpha\beta}(y)
\geq \half |\langle [F_{\mu\nu}(x) , F_{\alpha\beta}(y)]\rangle|
\label {4a}  \ee
If the points $x$ and $y$ are space-like separated, these measurements
should not interfere with each other --- otherwise, it would be possible to
transmit signals acausally. Therefore it is important that measurable
fields like $F_{\mu\nu}$ should obey {\ital microcausality conditions}
\be   [F_{\mu\nu}(x),F_{\alpha\beta}(y)] = 0\ ,\hspace{4mm} (x-y)^2 < 0
\label{4b}  \ee

As we shall see, the result for free fields satisfies Eq.\ (\ref{4b}),
and illustrates another general characteristic of field commutators ---
that they are {\ital singular\/} on the light cone $(x-y)^2 = 0$.\\[4mm]
{\ital 4.1 Free \vspace{1mm}Fields}

The free four-potential ${\cal A}_\mu$ given by Eq.\ (\ref{3s}) is a
linear combination of operators $\Down$ and $\Up$ with commutators
given by Eq.\ (\ref{3r}). Therefore the field commutators for
${\cal A}^\mu(x)$ are
\be  [{\cal A}^\mu(x),{\cal A}^\nu(y)]
  = \invint{k}\sum^2_{\lambda = 1} \pol \varepsilon^{\nu *}_{\fk \lambda}
      \, e^{-ik\cdot (x-y)}\  -\  \mbox{\rm h.c.}
%\label{4c}
\ee
According to Eq.\ (\ref{3e5}), the sum over polarizations is $-g^{\mu\nu}$
plus gauge terms depending on $k^\mu$ and $N^\mu$, so we find
\be  [{\cal A}^\mu(x),{\cal A}^\nu(y)]
  = -i g^{\mu\nu}\{\Delta^+(x-y) - \Delta^+(y-x)\}\ +\ \mbox{\rm gauge terms}
\label{4d}  \ee
where the notation
\be  \Delta^+(x) = -i\invint{k}\, e^{-ik\cdot x}\ ,
                                \hspace{5mm} k_0 = |{\bf k}|
\label{4e}  \ee
is standard \cite{SSS,IZ,MS}; the superscript $+$ refers to the
positive-energy projector $\theta(k_0)$ in Eq.\ (\ref{3qa}).

As written, the integral (\ref{4e}) is not obviously convergent --- it
oscillates for $|{\bf k}|$ large. A similar problem is encountered
for the one-dimensional Fourier transform of the Heaviside step function
\be  \theta(k) = \left\{
    \begin{array}{ll} 1 \hspace{3mm}& \mbox{if $k > 0$} \\
                      0 & \mbox{if $k < 0$} \end{array} \right.
%\label{4e1}
\ee
Oscillations in $\int\! dk\,\theta(k)e^{-ik\cdot x}$ at $k=+\infty$
are controlled by substituting $x\rightarrow x - i\epsilon$ for positive
$\epsilon$ and taking the limit $\epsilon \rightharpoonup 0$:
\be \int\!\! dk\,\theta(k)e^{-ik\cdot x}
  =  \begin{array}{c}\lim \\[-1.2mm] \epsilon \rightharpoonup 0 \end{array}
     \int^\infty_0\!\! dk\, e^{-ik(x-i\epsilon)}
  = -i \begin{array}{c}\lim \\[-1.2mm] \epsilon \rightharpoonup 0 \end{array}
                (x-i\epsilon) ^{-1}
\label{4e2}   \ee
Usually this is written $-i/(x-i\epsilon)$ with the limiting
procedure understood, as in the formula
\be (x-i\epsilon)^{-1} - (x+i\epsilon)^{-1} = 2\pi i\,\delta(x)
\label{4f}  \ee

Similarly, let the four-vector $x^\mu$ in Eq.\ (\ref{4e})
become complex:
\be  x^\mu \rightarrow x^\mu - i\eta^\mu\ , \ \ \eta^2 > 0\ , \ \eta_0 > 0
\label{4g}  \ee
The four-vector $\eta^\mu$ is restricted to lie within the forward light
cone to ensure convergence of the integral
\be  \Delta^+(x-i\eta) = -i\invint{k}\, e^{-ik\cdot(x-i\eta)}
\label{4g1} \ee
To calculate it, choose the frame $\eta^\mu = (\epsilon,0,0,0)$,
and introduce polar coordinates $(|{\bf k}|,\theta,\phi)$ for ${\bf k}$ with
the Z-axis along ${\bf x}$:
\be \Delta^+(x-i\eta) = -\frac{i}{8\pi^2}\int^\infty_0\!\! d|{\bf k}|\,
 |{\bf k}|\, e^{-i|\sk|(x_0-i\epsilon)}\int^1_{-1}\!\!
 d(\cos\theta) e^{i|\sk||\sx|\cos\theta}
= \frac{i}{4\pi^2(x-i\eta)^2}
\label{4g2}  \ee
Then the expression (\ref{4e}) is given by the limit $\eta^\mu \rightarrow 0$,
i.e.
\be  \Delta^+(x) = \frac{i}{4\pi^2(x^2 - i\epsilon x_0)}
\label{4g3} \ee
in conventional notation. Notice that $\del^2\Delta^+(x-i\eta)$ vanishes for
all $\eta$ within the forward light cone, so the same is true for the
boundary value (\ref{4g3}):
\be  \del^2\Delta^+(x) = 0   \label{4g4}  \ee

The commutator (\ref{4d}) involves the Jordan-Pauli function \cite{JP}
\be  \Delta(x) = \Delta^+(x) -\Delta^+(-x)  \label{4h} \ee
 From Eqs.\ (\ref{4f}) and (\ref{4g3}), we find
\be  \Delta(x) = \frac{i}{4\pi^2}\left\{\frac{1}{x^2 - i\epsilon x_0}
                               - \frac{1}{x^2 + i\epsilon x_0}\right\}
               = -\frac{1}{2\pi}\varepsilon(x_0)\delta(x^2)
\label{4i}  \ee
where $\varepsilon(t)$ takes values $\pm 1$ according to the sign of $t$:
\be \varepsilon(t) = t/|t| = \theta(t) - \theta(-t)  \ee

The gauge terms in (\ref{4d}) correspond to the formal substitution
$k_\mu \rightarrow i\del_\mu$ in the polarization sum (\ref{3e5}):
\be  [{\cal A}^\mu(x),{\cal A}^\nu(y)]
= i\left\{ - g^{\mu\nu} + (N^\mu\del^\nu + N^\nu\del^\mu)(N.\del)^{-1}
   - \del^\mu\del^\nu (N.\del)^{-2}\right\} \Delta(x-y)
\label{4j} \ee
Some caution is necessary here because the requirements
\be  N.\del\, (N.\del)^{-1}\ =\ 1\ =\ (N.\del)^{-1}N.\del   \ee
do not fix $(N.\del)^{-1}$ uniquely --- the pole $(N.k \pm i\epsilon)^{-1}$
in Fourier space can be specified in various ways.

For example, the notation $\del^{-1}_0$ could refer to either of the
prescriptions
\be  (\del_0\pm\epsilon)^{-1}f(t,{\bf x})
   =  \int^t_{\mp\infty}\!\! dt' f(t',{\bf x})  \ee
or a mixture thereof. However the ambiguity disappears if $f(t,{\bf x})$
can be written in the form
\be  f(t,{\bf x}) = F(t,{\bf x})  - F(t-c,{\bf x})\ ,\ \ c = {\rm constant}
\label{4k}  \ee
We find uniquely:
\be  \del^{-1}_0 f(t,{\bf x}) = \int^t_{t-c}\!\! dt'\,F(t',{\bf x})  \ee

The commutator function $\Delta(x)$ is of the form (\ref{4k}),
\be  \Delta(x) = \frac{1}{4\pi|{\bf x}|}
    \bigl(\delta(t+|{\bf x}|) - \delta(t-|{\bf x}|)\bigr)
\label{4l}  \ee
so we can calculate
\be  \del^{-1}_0 \Delta(x) = \frac{1}{4\pi|{\bf x}|}
    \bigl(\theta(t+|{\bf x}|) - \theta(t-|{\bf x}|)\bigr)   \ee
and
\be  \del^{-2}_0 \Delta(x) = \frac{1}{8\pi|{\bf x}|}
    \left(\bigl|t+|{\bf x}|\bigr| - \bigl|t-|{\bf x}|\bigr|\right)
\label{4m}  \ee
uniquely. A check is to calculate the inverse Laplacian (\ref{3b0}) of
$\Delta(x)$ and obtain
\be  \nabla^{-2} \Delta(x) = \del^{-2}_0 \Delta(x)   \ee
in agreement with the property $\del^2 \Delta(x) = 0$ implied by Eqs.\
(\ref{4g4}) and (\ref{4h}). As a result, the non-vanishing commutators
(\ref{4j}) for the case $N^\mu = (1,0,0,0)$ can be written
\be  [{\cal A}_i(x),{\cal A}_j(y)]
  = i\left(\delta_{ij} - \nabla_i\nabla_j\nabla^{-2}\right)\Delta(x-y)\ ,
  \hspace{5mm} i,j = 1,2,3
\label{4n}  \ee
The generalization of this discussion to other timelike vectors $N^\mu$
is tedious but straightforward.

The gauge terms in Eq.\ (\ref{4j}) are total derivatives, so it follows
from Eq.\ (\ref{3d}) that they do not contribute to $F_{\mu\nu}$ commutators.
Consequently, the result for free fields is \cite{JP}
\be [F_{\mu\nu}(x),F_{\alpha\beta}(y)] = \frac{i}{2\pi}
  \left( \del_\mu \delta^\sigma_\nu - \del_\nu \delta^\sigma_\mu \right)_x
 \left( \del_\alpha g_{\sigma\beta} - \del_\beta g_{\sigma\alpha} \right)_y
 \left\{\varepsilon(x_0 - y_0)\,\delta\!\left((x-y)^2\right)\right\}
\label{4o}  \ee
As promised, the result vanishes for space-like separations and is
singular on the light cone.

Eq.\ (\ref{4o}) has special features not characteristic of the interacting
case:
\begin{description}
\item[{\makebox[7mm][l]{\elevenrm (a)}}] It is conserved in each of the
indicies $\mu,\nu,\alpha,\beta$. This is required by the free-field
equations (\ref{3a}), but cannot be maintained in the presence of the
current operator $j_\nu$ in Maxwell's equations (\ref{2a}).
\item[{\makebox[7mm][l]{\elevenrm (b)}}] It is proportional to the unit
operator $I$.
\item[{\makebox[7mm][l]{\elevenrm (c)}}] It vanishes for time-like
separations. This property is peculiar to massless free fields.
\end{description}

Notice that the gauge terms in Eqs.\ (\ref{4j}) and (\ref{4n}) do
{\ital not\/} vanish at spacelike separations. This becomes clear when Eq.
(\ref{4m}) is written
\be \del^{-2}_0 \Delta(x) = \frac{t}{4\pi|{\bf x}|}\theta(-x^2)
        + \frac{1}{4\pi}\varepsilon(t)\theta(x^2)   \ee
Of course, these effects are not physical, so microcausality is not
in question.\\[4mm]
{\ital 4.2 Measuring Electric and Magnetic \vspace{1mm}Fields}

At first sight, the light-cone singularity of the commutator (\ref{4o})
seems to be a problem. If the right-hand side of Eq.\ (\ref{4a}) can be
infinite, are the components of $F_{\mu\nu}$ measurable in any sense?

Bohr and Rosenfeld \cite{Bohr} observed that the description of
${\bf E}$ and ${\bf B}$ as field components at each space-time point
is an idealization of the actual physical situation. Observed quantities
are really averages of these field components over various space-time
regions. In other words, one can discuss the observed values of $F_{\mu\nu}$
components for a {\ital neighbourhood} of any space-time point, but not
the value at such a  point.

In modern terminology \cite{SW,Jost,Haag}, the operators ${\bf E}(x)$ and
${\bf B}(x)$ are {\ital generalized functions\/}, or {\ital distributions\/}:
they represent the set of smeared operators
\be  F_{\mu\nu}[f] = \Int{4}{x}\, f(x) F_{\mu\nu}(x)  \label{4p}  \ee
corresponding to all smooth functions $f(x)$ decreasing rapidly at
$x_\mu \sim \infty$ (faster than any inverse power). A familiar classical
example of this is the charge density $\rho({\bf x})$ of a point charge
$Q$ at ${\bf a}$:
\be  \rho({\bf x}) = Q\,\delta^3({\bf x - a})   \label{4p1}  \ee
Eq.\ (\ref{4p1}) really refers to the linear functional $\rho$ which assigns
the number $Q\, g({\bf a})$ to each ``test'' function $g=g({\bf x})$:
\be  g({\bf x})\ \stackrel{\rho}{\rightarrow}\ Q\, g({\bf a}) = \rho[g]  \ee
A similar interpretation is to be given to the boundary values of
complex functions, such as Eqs.\ (\ref{4e2}) and (\ref{4g3}). Note that
smearing for $F_{\mu\nu}(x)$ and associated distributions like $\Delta(x)$
is over {\ital both\/} space and time.

Smearing becomes essential if we are to have a sensible interpretation of
the uncertainty $\Delta F_{\mu\nu}$. Without smearing,
$\Delta F_{\mu\nu}(x)$ would have to be represented algebraically as
follows,
\be  \sqrt{\langle  F_{\mu\nu}(x) F_{\mu\nu}(x)
        - \langle F_{\mu\nu}(x)\rangle^2\rangle}
%\label{4p2}
\ee
with no summation over $\mu,\nu$. The trouble with this formula is that
operators are multiplied together at the same space-time point $x_\mu$.
It is clear from the commutator (\ref{4o}) or the discussion of normal
ordering below Eq.\ (\ref{3m1}) that such an expression is infinite, by
construction! Instead, we should consider the uncertainty in $F_{\mu\nu}[f]$
for each test function $f$:
\be \Delta F_{\mu\nu}[f]
  =  \sqrt{\langle  F_{\mu\nu}[f]^2 - \langle F_{\mu\nu}[f]\rangle^2\rangle}
\ ,\ \ \ \mbox{\rm no sum over $\mu,\nu$}
%\label{4p3}
\ee
Then the uncertainty relations
\be \Delta F_{\mu\nu}[f]\, \Delta F_{\alpha\beta}[g]
\geq \half |\langle [F_{\mu\nu}[f] , F_{\alpha\beta}[g]]\rangle|
\label{4p4} \ee
make perfect sense when the free-field result (\ref{4o}) is substituted.

A less formal version of this conclusion is that infinite results will not
be observed because the light-cone singularity in Eq. (\ref{4o}) is
{\ital integrable\/}. This crucial observation was the starting point
for the work of Bohr and Rosenfeld \cite{Bohr}. After a long analysis
of carefully constructed thought experiments, they managed to show
how the constraints implied by Eqs.\ (\ref{4o}) and (\ref{4p4}) can
be deduced from the uncertainty relations
\be \left(\Delta p \Delta x\right)_{\mbox{\ninerm probe}}
    \geq \half\hbar  \ee
obeyed by any probe used to measure electromagnetic field strengths. The
discussion is not easily summarized, as several reviewers \cite{BRrev}
have discovered.

An important consequence of the Bohr-Rosenfeld analysis is that
a field such as $A_\mu$ coupled causally to quantized matter
cannot remain classical --- it must itself be quantized.
\sect{5}{Local Quantum Field Theories}

As we saw in Section 1, quantum fields are needed to describe all
relativistic matter, not just photons. So we consider a collection
of fields
\be \phi_j = \phi_j (x)
%\label{5a1}
\ee
The subscript $j$ refers to spin, charge or other distinguishing
quantum numbers carried by each field component.\\[4mm]
{\ital 5.1 General \vspace{1mm}Properties}

In local theories, all field components are assumed to depend on a
single\footnote{String theories \cite{strings} generalize local field
theory. They involve operators depending on a line of space-time points
(string). By Taylor expansion, each operator of this type can be
regarded as equivalent to an infinite set of local field operators.}
space-time variable $x^\mu$, such that active translations of states
through a space-time interval $a^\mu$
\be |\psi\rangle \longrightarrow |\psi_a\rangle = e^{-iP.a}|\psi\rangle
%\label{5a2}
\ee
correspond to a shift $x^\mu \rightarrow x^\mu + a^\mu$ in the
space-time dependence of these operators:
\be
\langle \xi_a|\phi_j(x+a)|\psi_a\rangle = \langle \xi|\phi_j(x)|\psi\rangle
\label{5a3}  \ee
Eq.\ (\ref{5a3}) holds for all states $|\psi\rangle$ and $|\xi\rangle$, so
the translation property previously noted for the free photon field in
Eq.\ (\ref{3p}) is obtained as a general result:
\be \phi_j(x+a) = e^{iP.a} \phi_j(x) e^{-iP.a}  \label{5a4} \ee

A similar property holds for Lorentz transformations
\be  x \longrightarrow x^\prime = \Lambda x
%\label{5a5}
\ee
and the corrresponding unitary operator $U_\Lambda$.
The new feature is that those indices $j$ of $\phi_j$ labelling spin
components are mixed:
\be  U_\Lambda \phi_j(x) U^\adj_\Lambda
              = \phi_i(\Lambda x){\cal D}_{ij}(\Lambda)
\label{5a6}  \ee
Matrices ${\cal D}(\Lambda)$ constructed from the coefficients
${\cal D}_{ij}(\Lambda)$ form a representation of the Lorentz group:
\be {\cal D}(\Lambda_1\Lambda_2) = {\cal D}(\Lambda_1){\cal D}(\Lambda_2)
%\label{5a7}
\ee

Another general property is microcausality, already considered for
the electromagnetic field strength in Eq.\ (\ref{4b}). Here it is
necessary to distinguish bosonic and fermionic field components.
Consider components $\phi_i[f]$ and $\phi_j[g]$ smeared over
functions $f(x)$ and $g(x)$ with space-like separated ``supports''
(regions in which the functions are non-zero almost everywhere).
According to Fermi-Dirac statistics, the state
\be  \ket{i,f;j,g;\psi} = \phi_i[f]\phi_j[g]\,\ket{\psi}
%\label{5a8}
\ee
should be anti-symmetric under the interchange
\[ \left\{\begin{array}{c}i\\f\end{array}\right\} \longleftrightarrow
         \left\{\begin{array}{c}j\\g\end{array}\right\}  \]
if both $\phi_i$ and $\phi_j$ are fermionic; otherwise, the interchange
should be symmetric\footnote{For different components $i\not= j$,
including the case of a field and its adjoint or a bosonic and a
fermionic component, the ``wrong'' symmetry may occur. However, such
a theory must then have the special property that new field components
with the ``right'' symmetry can be constructed from the old fields via
a ``Klein transformation'' \cite{SW}.}.
So the general form of the microcausality condition is
\be  [\phi_i(x), \phi_j(y)]_\mp = \phi_i(x)\phi_j(y) \mp \phi_j(y)\phi_i(x)
                                = 0\ , \ \ \ (x-y)^2 < 0
\label{5a9} \ee
where the anti-commutator $[\ ,\; ]_+$ is used only if both $\phi_1$ and
$\phi_2$ are fermionic. Note that the anti-commutator version of (\ref{5a9})
does not contradict the independence of measurements at space-like
separations because such measurements are possible only for {\ital even\/}
powers of fermionic operators.

The book of Streater and Wightman \cite{SW} analyses these axiomatic
requirements in a manner which is advanced and rigorous but nevertheless
instructive. The most important results of axiomatic field theory are that:
\begin{description}
\item[{\makebox[7mm][l]{\elevenrm (a)}}] Field components $\phi_i$
are bosonic or fermionic according to whether the Lorentz representation
${\cal D}(\Lambda)$ in Eq.\ (\ref{5a6}) is tensor (integer spin)
or spinor (half-integer spin) respectively. This is the spin--statistics
theorem, first obtained for free fields by Fierz \cite{Fierz} and Pauli
\cite{Pauli2}; the general proof \cite{spinstat} followed much later.
\item[{\makebox[7mm][l]{\elevenrm (b)}}] The product $PCT$ of parity,
charge conjugation and time reversal is always a symmetry of a local
quantum field theory. The $PCT$ theorem was first stated in this general
form by Pauli \cite{Pauli3}.
\end{description}
\vspace*{4mm}
{\ital 5.2 Free \vspace{1mm}Fields}

The arguments leading to the formula (\ref{3s}) for the free electromagnetic
four-potential can be readily carried over to other types of field. Indeed,
the analysis is simpler for spin-0 and spin-$\half$ fields because there is
no gauge-fixing.

Instead of Maxwell's theory, we consider a general Lagrangian (density)
with field dependence
\be {\cal L}
= {\cal L}(\phi_1,\del\phi_1, \ldots \phi_j,\del\phi_j,\ldots )
\label{5b1}  \ee
For complex components, the field and its conjugate are treated as separate
variables in Eq.\ (\ref{5b1}). According to classical field theory, this
Lagrangian corresponds to the equations of motion
\be  \frac{\del{\cal L}}{\del\phi_j}
    = \del_\mu \frac{\del{\cal L}}{\del\del_\mu\phi_j}
\label{5b2}  \ee
and the Hamiltonian (density)
\be  {\cal H}
   = \sum_j\frac{\del{\cal L}}{\del\del_0\phi_j}\del_0\phi_j - {\cal L}
%\label{5b3}
\ee

Free fields satisfy linear equations such as the equation for a free scalar
field $\phi$,
\be  \left(\del^2 + m^2\right)\phi = 0
%\label{5b4}
\ee
so the Lagrangian is a quadratic form, e.g.
\be  {\cal L} = \half \del_\mu\phi\del^\mu\phi - \half m^2 \phi^2 \ ,\ \
{\rm real}\ \phi
\label{5b5}  \ee
or
\be  {\cal L} = \del_\mu\phi^\adj \del^\mu\phi - m^2 \phi^\adj \phi \ ,\ \
{\rm complex}\ \phi
\label{5b6}  \ee
Evidently, there are no polarization factors in the analogue of Eq.
(\ref{3s}) for a spin-0 field:
\be
\phi(x) = \invint{k} \left\{ a({\bf k}) e^{-ik\cdot x}
                + b^\adj({\bf k}) e^{ik\cdot x}\right\}\ ,\ \
                   k_0 = \sqrt{|{\bf k}|^2 +m^2}
\label{5b7}  \ee
If $\phi$ is real, the creation operator $b^\adj({\bf k})$ is just
$a^\adj({\bf k})$, as in Eq.\ (\ref{3s}). Otherwise, $b^\adj({\bf k})$
creates particles which are distinct from those created by
$a^\adj({\bf k})$; indeed they are anti-particles having the same mass
but opposite charge. Thus complex fields are required when particles
and their anti-particles are distinct --- unlike photons, which are their
own anti-particles.

The normalizations of the Lagrangians (\ref{5b5}) and (\ref{5b6}) are
fixed by convention. Then an argument similar to that leading to
Eqs.\ (\ref{3n}) and (\ref{3s}) determines constant factors in
Eq.\ (\ref{5b7}) by requiring $\Int{3}{x}:{\cal H}(x):$ to be the
energy operator, with normal ordering as in Eq.\ (\ref{3m1}).

The commutators of the free scalar field $\phi$ involve the massive version
of the invariant function $\Delta^+(x)$ discussed in Section 4.1,
\be  \Delta^+(x\, ;m)
= -i\begin{array}{c}\lim \\[-1.2mm] \eta \rightharpoonup 0 \end{array}
    \invint{k}\, e^{-ik\cdot(x-i\eta)}\ ,\ \ k_0 = \sqrt{|{\bf k}|^2 +m^2}
\label{5b8}  \ee
where the symbol $\rightharpoonup$ indicates that the limit is taken
through positive time-like values of $\eta$.
If $\phi$ is real, direct calculation using Eq.\ (\ref{3r}) yields
\be [\phi(x),\phi(y)] = i\Delta(x-y\, ;m)
%\label{5b9}
\ee
where the massive Pauli-Jordan function \cite{SSS}
\bea \Delta(x\, ;m) &=& \Delta^+(x\, ;m) - \Delta^+(-x\, ;m) \nonumber \\
       &=& -\frac{1}{2\pi}\varepsilon(x_0)\left\{\delta(x^2)
     - \frac{m}{2\sqrt{x^2}}\theta(x^2)J_1\!\left(m\sqrt{x^2}\right)\right\}
\label{5c2}  \eea
contains a Bessel function $J_\nu(z)$ of order $\nu = 1$.
When $\phi$ is complex, $a({\bf k})$ and $b({\bf k})$ and their conjugates
separately obey Eq.\ (\ref{3r}) but otherwise commute, so we find
\be  [\phi(x),\phi^\adj(y)] = i\Delta(x-y\, ;m)\ , \ \
     [\phi(x),\phi(y)] = 0
\label{5c3}  \ee
It is evident from the explicit expression (\ref{5c2}) that these
commutators exhibit microcausality. Notice that $\Delta^+(x\, ;m)$ and hence
$\Delta(x\, ;m)$ are annihilated by the massive wave operator
$\left(\del^2 + m^2\right)$.

The free spin-$\half$ field $\psi(x)$ satisfies the Dirac
equation\footnote{Relativistic quantum mechanics is covered in
the first volume of Bjorken and Drell \cite{BjD1} and in Chapter 2 of
Itzykson and Zuber \cite{IZ}.}
\be  (i\!\!\not\!\del - m)\psi(x) = 0  \label{5c4}  \ee
where Feynman's ``slash notation''
\be  \not\!\!{A} = \gamma^\mu A_\mu   \label{5c5}  \ee
is used for the first-order differential operator\ $\not\hspace{-1mm}\del =
\gamma^\mu \del_\mu$\,, and the $4\times4$
matrices $\gamma^\mu$ satisfy the Clifford algebra
\be  [\gamma_\mu , \gamma_\nu ]_+ = 2 g_{\mu\nu}  \label{5c6}  \ee
Note that the matrix operator in (\ref{5c4}) is really
$i\!\!\not\hspace{-1mm}\del - mI$, where $I$ is the unit $4\times 4$ matrix.
Similarly, the right-hand side of Eq.\ (\ref{5c6}) is $2g_{\mu\nu}I$.

Usually, $\psi$ is associated with particles such as electrons which
are distinct from their anti-particles (positrons), so its Fourier
decomposition is of the form
\be
\psi(x) = \sum_s\invint{p} \left\{ c_s({\bf p})u_s({\bf p}) e^{-ip\cdot x}
                + d^\adj_s({\bf p})v_s({\bf p}) e^{ip\cdot x}\right\}\ ,\ \
                   p_0 = \sqrt{|{\bf p}|^2 + m^2}
\label{5c7}  \ee
where the operators $c_s({\bf p})$ and $d_s({\bf p})$ annihilate (say)
electrons and positrons of spin projection $s$ (up or down in the rest
frame), and the corresponding four-spinors $u_s({\bf p})$ and $v_s({\bf p})$
satisfy momentum-space Dirac equations
\be  (\!\not\!p - m)u_s({\bf p}) = 0\ , \ \
      (\!\not\!p + m)v_s({\bf p}) = 0\
\label{5c8}  \ee
In terms of inner products taken between spinors and their adjoints
\be   \overline{\rm spinor} = \{{\rm spinor}\}^\adj \gamma^0
%\label{5c9}
\ee
the momentum-space spinors $u_s({\bf p}),v_s({\bf p})$ are mutually orthogonal
\be   \bar{u}_s({\bf p})u_{s^\prime}({\bf p}) = 2m\delta_{s s^\prime}
      = -\bar{v}_s({\bf p})v_{s^\prime}({\bf p})\ , \ \
      \bar{u}_s({\bf p})v_{s^\prime}({\bf p}) = 0
\label{5d1}  \ee
in a convenient normalization. Therefore they satisfy the following
completeness relation in four-spinor space:
\be  \sum_s \{ u_s({\bf p})\bar{u}_s({\bf p})
                      - v_s({\bf p})\bar{v}_s({\bf p}) \}  = 2m
\label{5d2}  \ee
To separate Eq.\ (\ref{5d2}) into projectors for positive and negative energy
solutions, left-multiply by the matrices $\,\not\!p \pm m$ and apply
Eq.\ (\ref{5c8}):
\be  \sum_s  u_s({\bf p})\bar{u}_s({\bf p})  =  \,\not\!p + m\ ,\ \
     \sum_s  v_s({\bf p})\bar{v}_s({\bf p})  =  \,\not\!p - m
\label{5d3}  \ee
{\ital Remarks\/}:
\begin{description}
\item[{\makebox[7mm][l]{\elevenrm (a)}}] If $\psi$ is massless ($m=0$),
the label $s$ refers to helicity, and the conditions
\be   u^\adj_s({\bf p})u_{s^\prime}({\bf p}) = 2p_0\delta_{s s^\prime}
      =  v^\adj_s({\bf p})v_{s^\prime}({\bf p})
%\label{5d4}
\ee
can be used instead of Eq.\ (\ref{5d1}). Experimental evidence is
consistent with neutrinos and anti-neutrinos having no mass.
\item[{\makebox[7mm][l]{\elevenrm (b)}}] Theoretically it is
possible for a fermion to be its own anti-fermion. Then the operators
$c_s({\bf p})$ and $d_s({\bf p})$ are identical and $\psi$ is called a
{\ital Majorana field\/}.
\end{description}

Given the normalizations used in Eqs.\ (\ref{5c7}) and (\ref{5d1}),
the  Lagrangian for Eq.\ (\ref{5c4}) can be written
\be  {\cal L} = \bar{\psi}(i\!\!\not\!\del - m)\psi   \label{5d5}  \ee
An alternative, which treats $\psi$ and $\bar{\psi}$ symmetrically, is
\be  \widetilde{\cal L} = \bar{\psi}({\textstyle \frac{i}{2}}\!\!\not\!
        \del^{\hspace{-2.5mm}\raisebox{1mm}{$\leftrightarrow$}} - m)\psi
\label{5d6}  \ee
where $\,\del^{\hspace{-2.5mm}\raisebox{1mm}{$\leftrightarrow$}}_\mu$ is
defined by the formula
\be   A\,\del^{\hspace{-2.5mm}\raisebox{1mm}{$\leftrightarrow$}}_\mu B
    \equiv  A\left(\del_\mu B\right) - \left(\del_\mu A\right)B
%\label{5d7}
\ee
The difference between Eqs.\ (\ref{5d5}) and (\ref{5d6}) is a
four-divergence $\half \del_\mu\{ \bar{\psi}\gamma^\mu\psi\}$ which
does not contribute to the equation of motion (\ref{5b2}). Both
Lagrangians contain an extra factor $\half$ if the field is Majorana,
as in the Lagrangian (\ref{5b5}) for a real scalar field.

The operators $c_s({\bf p})$ and $d_s({\bf p})$ destroy fermions, so
anti-commutator relations \cite{JW} are necessary:
\[ [c_s({\bf p}),c^\adj_{s^\prime}({\bf q})]_+
 =  [d_s({\bf p}),d^\adj_{s^\prime}({\bf q})]_+
 =  2p_0 \delta_{s s^\prime}(2\pi)^3\delta^3({\bf p}-{\bf q})\ , \]
\be [c_s({\bf p}),c_{s^\prime}({\bf q})]_+
 =  [c_s({\bf p}),d_{s^\prime}({\bf q})]_+
 =  [c_s({\bf p}),d^\adj_{s^\prime}({\bf q})]_+
 =  [d_s({\bf p}),d_{s^\prime}({\bf q})]_+ = 0
\label{5d8}  \ee
Indeed, when Eq.\ (\ref{5c7}) is substituted into the Hamiltonian
derived from ${\cal L}$ or $\widetilde{\cal L}$, normal ordering must
include a minus sign for each fermion interchange
\be  :c_s({\bf p})c^\adj_{s^\prime}({\bf q}):
 = - :c^\adj_{s^\prime}({\bf q})c_s({\bf p}):\ ,\ \
     :d_s({\bf p})d^\adj_{s^\prime}({\bf q}):
 = - :d^\adj_{s^\prime}({\bf q})d_s({\bf p}):
%\label{5d9}
\ee
if the desired result
\be \Int{3}{x}\, :{\cal H}:\ = \invint{p}
    \{c^\adj_s({\bf p})c_s({\bf p}) + d^\adj_s({\bf p})d_s({\bf p})\} p_0
\label{5e1}  \ee
is to be obtained. Similarly, the field anti-commutator can be calculated
from Eqs.\ (\ref{5c7}), (\ref{5d3}) and (\ref{5d8}) to obtain a result
\be [\psi_\sigma(x),\bar{\psi}_\tau(y)]_+
  = (i\!\!\not\!\del + m)_{\sigma\tau}\, i\Delta(x-y\, ;m)
\label{5e2}  \ee
consistent with microcausality. As one might expect from the
spin-statistics theorem, consistent results for $\psi$ are not possible
if one tries to use commutators throughout. \\[4mm]
{\ital 5.3 Canonical \vspace{1mm}Methods}

We must now consider how to quantize interacting field theories such
that explicit calculations can be performed. The most popular procedures
are:
\begin{description}
\item[{\makebox[7mm][l]{\elevenrm (a)}}] Canonical quantization, pioneered
by Heisenberg and Pauli \cite{HP};
\item[{\makebox[7mm][l]{\elevenrm (b)}}] Functional methods, particularly
Feynman's path-integral approach \cite{Feynman}.
\end{description}
Generally, calculations are possible only if one expands in the strength
of the interaction, in successive orders of perturbation theory.

An example of an interacting theory is the scalar-field Lagrangian
(\ref{5b5}) with an extra term $-\lambda\phi^4$,
\be  {\cal L}
= \half \del_\mu\phi\del^\mu\phi - \half m^2 \phi^2 - \lambda\phi^4
\label{5e3}  \ee
where the real coupling constant $\lambda$ fixes the strength of the
interaction. If so desired, another interaction could be included by
adding the free-fermion Lagrangian (\ref{5d5}) and a ``scalar Yukawa
coupling'' $G\bar{\psi}\psi\phi$ to Eq.\ (\ref{5e3}),
\be  {\cal L}
= \half \del_\mu\phi\del^\mu\phi - \half m^2 \phi^2 - \lambda\phi^4
 + \bar{\psi}(i\!\!\not\!\del - m + G\phi)\psi
\label{5e4}
\ee
where $G$ is (in this case) also real. Both $\lambda$ and $G$ become
expansion parameters for the perturbative expansion.

A more practical example is quantum electrodynamics\footnote{Both QED
and terms from (\ref{5e4}) are contained in the electroweak sector of
the Standard Model \cite{Lee,Ryder,IZ,MS} which is now the basis for
particle physics phenomenology \cite{data}.} (QED),
obtained from (\ref{5d5}) by making the derivative covariant
\be  \del_\mu \longrightarrow D_\mu = \del_\mu - ieA_\mu
%\label{5e5}
\ee
and adding the Lagrangian $-{\textstyle \frac{1}{4}}F^2$ for pure
Maxwell theory:
\be  {\cal L}
 = - {\textstyle \frac{1}{4}}F_{\mu\nu}F^{\mu\nu}
   + \bar{\psi}\left(i\!\!\not\!\!D - m\right)\psi
%\label{5e6}
\ee
Here $-e$ is the ``bare'' electronic charge. The measured or
``renormalized'' charge $-1.6\times10^{-19}$ Coulomb corresponds
to $-e$ plus perturbative corrections due to self-interactions of
the electron.

Canonical theory treats fields $\phi_j(x)$ as dynamical variables
evaluated at an {\ital instant\/} of time. Three-dimensional smearing
\be  \phi_{j,t}[g] = \Int{3}{x} g({\bf x}) \phi_j(t,{\bf x})
%\label{5e7}
\ee
is assumed to be adequate --- time $t$ is relegated to the status of a
continuous label. Distributions with this property are said to be ``sharp''
in time.

The procedure resembles that of ordinary quantum mechanics, where
independent dynamical variables $q_m(t)$ and their conjugate momenta
$p_m(t)$ are required to satisfy
\be  [q_m(t),p_n(t)] = i\delta_{mn}\ ,\ \
     [q_m(t),q_n(t)] = 0 = [p_m(t),p_n(t)]
\label{5e8}  \ee
The only difference is that the Kronecker delta has to be generalized
to include a three-dimensional delta function. So fields
$\phi_j(t,{\bf x})$ and their ``generalized momenta''
\be  \pi_j(t,{\bf x}) = \frac{\del{\cal L}}{\del\phi_j(t,{\bf x})}
%\label{5e9}
\ee
are postulated to obey the equal-time relations\footnote{This assumes that
there are no constraints on the dynamical variables. The general theory
for constrained systems was developed by Dirac \cite{Dirac2,IZ}. In
quantum electrodynamics and other gauge theories, this is associated with
the need to fix a gauge.} \cite{HP}
\be [\phi_j(t,{\bf x}),\pi_k(t,{\bf y})]_\mp
                            = i\delta_{jk}\delta^3({\bf x-y})
\label{5f1} \ee
\be [\phi_j(t,{\bf x}),\phi_k(t,{\bf y})]_\mp = 0
                            = [\pi_j(t,{\bf x}),\pi_k(t,{\bf y})]_\mp
\label{5f2}  \ee
As usual, anticommutators are required if both $\phi_j$ and $\phi_k$
are fermionic.

The canonical hypotheses (\ref{5f1}) and (\ref{5f2}) can be tested
\cite{HP} by taking the equal-time limit of known free-field
(anti-)commutators such as (\ref{5c3}) and (\ref{5e2}). These involve
the invariant function $\Delta(x\, ;m)$ of Eq.\ (\ref{5c2}), whose
singular part at short distances $x_\mu \rightarrow 0$ is given
entirely by the mass-independent function $\Delta(x)$ of Eq.\ (\ref{4i}):
\be \Delta(x\, ;m)
     = -\frac{1}{2\pi}\varepsilon(x_0)\delta(x^2)
   + \frac{m^2}{4\pi}\varepsilon(x_0)\theta(x^2)\{1 + O(m^2x^2)\}\ ,
\ \ x_\mu \rightarrow 0   \label{5f3}  \ee
Let us apply the smearing operation $\Int{3}{x}g({\bf x})\ldots$ to
this expansion, and consider the region
$t\sim 0$. The mass-dependent part is finite and confined to the
region $|{\bf x}| \leq |t|$, so its asymptotic contribution is
\be \frac{m^2}{4\pi}\varepsilon(x_0)\int^{|t|}_0\! d|{\bf x}|\,
      |{\bf x}|^2\int\! d\Omega\, g({\bf x})
  \sim {\textstyle \frac{1}{3}}m^2 g(0)\, t^3\ ,\ \  t \rightarrow 0
%\label{5f4}
\ee
where $d\Omega$ is the element of solid angle in the direction
${\bf \hat{x}}$.
Therefore, we have
\bea   \Int{3}{x}\,\Delta(t,{\bf x}\, ;m)g({\bf x})
    &=& \Int{3}{x}\,\Delta(t,{\bf x})g({\bf x}) + O(t^3) \nonumber \\
    &=& - \frac{t}{4\pi}\int\!\! d\Omega\, g(|t|{\bf \hat{x}}) + O(t^3)
%\label{5f5}
\eea
where the last line follows from Eq.\ (\ref{4l}) for $\Delta(x)$. Then
we can substitute
\be g(|t|{\bf \hat{x}}) = \{1+|t|\,{\bf \hat{x}}\cdot\nabla + O(t^2)\}\, g(0)
%\label{5f6}
\ee
and note that the angular integral $\int\!\! d\Omega\, {\bf \hat{x}}$
vanishes:
\be \Int{3}{x}\,\Delta(t,{\bf x}\, ;m)g({\bf x}) = - t\, g(0) + O(t^3)
\label{5f7}  \ee
The relevant equal-time relations can be read from Eq.\ (\ref{5f7}).
There are no $O(t^0)$ or $O(t^2)$ terms,
\be \begin{array}{c}\lim \\[-1.2mm] t \rightarrow 0 \end{array}
  \Delta(t,{\bf x}\, ;m) = 0 =
    \begin{array}{c}\lim \\[-1.2mm] t \rightarrow 0 \end{array}
  \del^2_0\Delta(t,{\bf x}\, ;m)
\label{5f8}  \ee
but there is a term $O(t)$ which gives rise to the key result \cite{HP}:
\be \begin{array}{c}\lim \\[-1.2mm] t \rightarrow 0 \end{array}
   \del_0\Delta(t,{\bf x}\, ;m) = - \delta^3({\bf x})
\label{5f9}  \ee
Then it is evident that the postulates (\ref{5f1}) and (\ref{5f2})
work for free fields. For example:
\begin{description}
\item[{\makebox[7mm][l]{\elevenrm (a)}}] The Lagrangian (\ref{5b6})
for a free complex scalar field $\phi$ requires $\del_0\phi^\adj$ to be
conjugate to $\phi$. From the commutators (\ref{5c3}) and Eqs.\
(\ref{5f8}) and (\ref{5f9}), we find
\be [\phi(t,{\bf x}), \del_0\phi^\adj(t,{\bf y})] = i\,\delta^3({\bf x-y})
\label{5g1}  \ee
with other possibilities
$[\phi,\phi]$, $ [\phi,\phi^\adj]$, $[\del_0\phi,\del_0\phi^\adj]$, and
$[\del_0\phi^\adj,\del_0\phi^\adj]$ vanishing at equal times.
\item[{\makebox[7mm][l]{\elevenrm (b)}}] At equal times, the
free-fermion anti-commutator (\ref{5e2}) reduces to
\be [\psi_\sigma(t,{\bf x}),\bar{\psi}_\tau(t,{\bf y})]_+
  =  \left(\gamma^0\right)_{\sigma\tau}\delta^3({\bf x-y})
%\label{5g2}
\ee
which is consistent with $i\psi^\adj$ being conjugate to $\psi$.
\end{description}

In textbooks, Eq.\ (\ref{5f9}) is often obtained by combining Eq.\
(\ref{5b8}) for $\Delta^+(x\, ;m)$ with the first equality in Eq.\
(\ref{5c2}) and neglecting the limiting procedure
$\eta_\mu\rightharpoonup 0$:
\be \begin{array}{c}\lim \\[-1.2mm] t \rightarrow 0 \end{array}
   \del_0\Delta(t,{\bf x}\, ;m)
 = -i\invint{k}\,\del_0\left\{e^{-ik\cdot x} - e^{ik\cdot x}\right\}_{t=0}
%\label{5g3}
\ee
However this obscures the role of short-distance behaviour in determining
the result at equal times.

The derivation leading to Eqs.\ (\ref{5f8}) and (\ref{5f9})
illustrates a general rule \cite{Wilson} governing equal-time limits.
Non-zero contributions to {\ital any\/} causal (anti-)commutator
\be [\phi_j(x_0,{\bf x}),\phi_k(y_0,{\bf y})]_\mp
\label{5g4}  \ee
are restricted to the region
\be  |{\bf x-y}| \leq |x_0 - y_0|
%\label{5g5}
\ee
Therefore all equal-time commutators of the operators $\phi_j$ and of
their derivatives are {\ital entirely\/} determined by the asymptotic
behaviour of (\ref{5g4}) at short distances
\be  (x-y)_\mu \sim 0
%\label{5g6}
\ee

This observation is critical in determining the status of the canonical
commutation relations (\ref{5f1}) and (\ref{5f2}). The aim of these
postulates is to state general quantization conditions for {\ital
interacting\/} theories. If the postulates are {\ital really\/} valid,
the short-distance behaviour of the theory should not depend on the
interaction: it should be the {\ital same} as for free fields. It turns
out that this requirement {\ital is\/} satisfied in the lowest-order
``tree'' approximation (sec.\ 6.3 below), but generally {\ital not\/}
in higher orders where self-interactions occur.

This contradiction of the canonical postulates is caused by {\ital
renormalization\/}, a ``patch-up'' procedure used to absorb infinite
self-interaction ambiguities into physical constants such as charges
and masses. Despite its {\ital ad hoc\/} appearance, renormalization
produces a theory which satisfies general physical requirements such
as causality, unitarity of the $S$-matrix, and above all, agreement
with experiment. Originally it was invented as the final step in the
evolution of quantum electrodynamics from Dirac's 1927 theory \cite{Dirac},
with matter represented by non-relativistic wave functions, to the
complete relativistic field theory of Schwinger, Feynman, Tomonaga
and Dyson, in which all divergences are eliminated from
perturbative calculations. (The history of these developments can be
traced in the well-known collection of classic papers selected by
Schwinger \cite{Schwinger}.) Space and time do not allow me to cover
renormalization in any detail, so readers are referred to general
textbooks [4--10] and the book by Collins \cite{Collins}. However the
general idea is as follows.

Perturbative corrections to amplitudes such as the strength of an interaction
depend on the values of incoming and outgoing momenta. Therefore, if we
are to identify a coupling ``constant'' as a measurable parameter, we
have to specify a point in momentum space to which it refers. This
introduces into the theory a new scale $\mu$ with dimensions of mass:
amplitudes depend on $\mu$ as well as momenta and masses.

At short distances, the dependence on masses disappears, as in Eq.\
(\ref{5f3}), but the $\mu$-dependence survives. This is because
infinite ambiguities due to self interactions are ultra-violet
problems which persist even in the massless case. Generally, each new
order of perturbation theory introduces an extra logarithmic factor
$\ln(\mu^2 x^2)$ at short distances. At a given perturbative order, the
result for scalar fields is
\be [\phi_j(x),\pi_k(y)] \sim C\del_0\, {\rm Im}\,
 \frac{\ln^p\!\left\{\mu^2\left(-(x-y)^2+i\epsilon(x_0 - y_0)\right)\right\}}
      {(x-y)^2 - i\epsilon(x_0 - y_0)}\ ,\ \ x_\mu - y_\mu \rightarrow 0
\label{5g7} \ee
where $p$ is a positive integer, $C$ is a calculable constant, and\
\,${\rm Im}$\,\ denotes the imaginary part. In the corresponding
anti-commutator for spin-$\half$ fields, $\del_0$ is replaced by
$\not\!\del\gamma_0$.

As explained above, the short-distance result (\ref{5g7}) enables us to
{\ital calculate\/} what happens at equal times $x_0 \sim y_0$:
\be [\phi_j(x_0,{\bf x}),\pi_k(y_0,{\bf y})]_\mp
 \sim 2^{p+1}\pi^2 C \ln^p\!\left(\mu|x_0 - y_0|\right)\delta^3({\bf x-y})
\label{5g8}  \ee
Evidently the result is an asymptotic series in powers of
$\ln\left(\mu|x_0 - y_0|\right)$. It does {\ital not\/} converge in
the limit $x_0 - y_0 \rightarrow 0$. Therefore the canonical postulates
(\ref{5f1}) and (\ref{5f2}) {\ital must be abandoned\/}.

This crucial point was first recognised by axiomatic theorists, as
reported on p.~101 of Streater and Wightman \cite{SW}. It entered the
mainstream literature twenty-five years ago with the appearance of
Wilson's well-known work on operator products at short distances
\cite{Wilson,Collins}. Nevertheless, textbooks are still appearing with
the reverential label ``fundamental'' attached to the canonical
commutators (\ref{5f1}) and (\ref{5f2}). Alternatively, authors try to
incorporate renormalization into an equal-time framework by introducing
a cutoff $\Lambda$ used at intermediate stages of calculations to
control infinities and writing
\be [\phi_j(t,{\bf x}),\pi_k(t,{\bf y})]_\mp
       \stackrel{?}{=} iZ(\Lambda)\delta_{jk}\delta^3({\bf x-y})\ ,
       \ \Lambda \sim \infty
%\label{5g9}
\ee
where the constant $Z(\Lambda)$ is a power series in $\ln(\Lambda/\mu)$.
This is at best misleading; the fact is that, in perturbation
theory, equal-time limits diverge as powers of
$\ln\left(\mu|x_0 - y_0|\right)$.

The result (\ref{5g8}) shows that the operators $\phi_j(x)$ and $\pi_j(x)$
are not sharp in time. Each operator smeared over a space-time region,
no matter how small, is partly time-like relative to its immediate
neighbours, so it can be causally affected by them. Consequently, it
cannot be assumed that the fixed-$t$ dynamical variables
$\left\{\phi_j(t,{\bf x})\right\}$ are entirely independent of each other,
even though microcausality is obeyed exactly. This has
thwarted attempts to quantize fields by analogy with the standard
prescription (\ref{5e8}) for quantum mechanics. It is not clear that
such a prescription is either possible or necessary, but it would
be an interesting development if someone could provide a suitable
construction. In the early eighties, Symanzik \cite{Symanzik} considered
formulating quantum field theory in the Schr\"{o}dinger picture, but
his line has not been pursued since then.

Other canonical procedures are also vitiated by renormalization. For
example, normal ordering of composite operators such as the
Hamiltonians (\ref{3m1}) and (\ref{5e1}) is no longer adequate --- in
higher orders, additional operator subtractions become necessary. The
classical Noether construction for the current $j^\mu$ associated with
a transformation $\delta\phi_j$ on the fields $\phi_j$ is particularly
vulnerable, because it requires us to multiply two operators at the
same point:
\be j^\mu = \sum_j \frac{\del{\cal L}}{\del\del_\mu\phi_j}\delta\phi_j
%\label{5h1}
\ee
After renormalization, there is no guarantee that the divergence
$\del_\mu j^\mu$ will be given by $\delta{\cal L}$. The axial-vector
anomaly \cite{anomaly} is the most famous example of a Noether construction
breaking down.

Whenever renormalization produces a result at variance with canonical
expectations, the effect is said to be ``anomalous''. Anomalies have
occupied a central place in field-theoretic research during the past
twenty-five years, with direct applications to phenomenology,
especially in quantum chromodynamics \cite{QCD,Yndurain}, the gauge
theory of quarks and gluons. Consequently, I have taken some trouble
to emphasize the incompatibility of renormalization and canonical
theory.

I should conclude this criticism of canonical procedures by noting that
similar problems are encountered with functional methods. As we shall see,
the latter rely on the assumption that unsmeared fields which are space-like
separated or Euclidean are dynamically independent, so that they can
be diagonalized simultaneously.\\[4mm]
{\ital 5.4 Continuation to Euclidean \vspace{1mm}Space}

Local quantum field theories have one other general property worth noting:
it is always possible to continue amplitudes analytically in $x_\mu$
away from Minkowski space-time with  invariant interval
\be  x^\mu x_\mu = x^2_0 - {\bf x}^2
%\label{5h2}
\ee
to Euclidean four-space:
\be  x_\mu x_\mu = {\bf x}^2 + x^2_4
%\label{5h3}
\ee

In any quantum theory, there must be a state of lowest energy, the
``ground state''. In a quantum field theory, the ground state
is known as the vacuum $\vac$. Generally, it lacks any particle
structure or other physical characteristics attached to a particular
frame of reference. Thus the property which distinguishes $\vac$
from other state vectors is its Poincar\'{e} invariance:
\be  P_\mu\vac = 0\ ,\ \ U_\Lambda\vac = 0
\label{5h4}  \ee

An example of such a state is the no-photon state $\ket{0}$ in
free-photon theory. The requirement that it have zero energy led
to normal ordering for the photon Hamiltonian in Eqs.\ (\ref{3m0})
and (\ref{3m1}). Free-field vacua satisfy special conditions of the
form
\be \down\ket{0} = 0
\label{5h5}  \ee
Generally Eq.\ (\ref{5h5}) is not applicable if there are interactions,
but Eq. (\ref{5h4}) retains its validity, irrespective of the properties
of other states in the theory.

Given $\vac$, we can construct vacuum expectation values
\be {\cal W} = \Vac\phi_1(x_1)\phi_2(x_2)\ldots\phi_n(x_n)\vac
\label{5h6}  \ee
known as ``unordered'' or ``Wightman'' functions. According to the
translation property (\ref{5a4}), the dependence on the space-time
points $(x_\mu)_j$ can be extracted in the following way:
\be  \phi_j(x_j) = e^{iP\cdot x_j}\phi_j(0)e^{-iP\cdot x_j}
\label{5h7}  \ee
Eq.\ (\ref{5h4}) requires the vacuum state to be translation invariant,
\be e^{-iP\cdot x}\vac = \vac
\label{5h8}  \ee
so ${\cal W}$ depends on coordinate differences only:
\be {\cal W} = {\cal W}(x_1-x_2,x_2-x_3,\ldots ,x_{n-1}-x_n)
\label{5h9}  \ee

A simple example is the two-point function for the scalar field (\ref{5b7}):
\be {\cal W}(x-y) = \bra{0}\phi(x)\phi^\adj(y)\ket{0}
                  = i\Delta^+(x\, ;m)
\label{5i1}  \ee
According to the definition (\ref{5b8}) of $\Delta^+$, the result is a
boundary value
\be {\cal W}(x-y)
 = \begin{array}{c}\lim \\[-1.2mm] \eta \rightharpoonup 0 \end{array}
   i\Delta^+(z\, ;m)
%\label{5i2}
\ee
where
\be  z_\mu = x_\mu - y_\mu - i\eta_\mu
%\label{5i3}
\ee
is a complex four-vector and $\eta_\mu$ lies within the forward light cone.
The case $m=0$ is discussed in detail in Eqs.\ (\ref{4g}) to (\ref{4g3}).

This connection with complex functions is a general property of vacuum
expectation values \cite{SW,Wightman}. It is derived from the fact that
all eigenvalues $p_\mu$ of the four-momentum operator $P_\mu$ must lie
on or within the forward light cone:
\be  p^2 \geq 0\ , \ \ p_0 \geq 0  \ee

To see this, insert sets of complete states $\ket{I}$ between the operators
$\phi_i$ in (\ref{5h6}),
\be {\cal W} = \sum_{I_1\ldots I_{n-1}}\Vac\phi_1(x_1)\ket{I_1}
   \bra{I_1}\phi_2(x_2)\ket{I_2}\ldots\bra{I_{n-1}}\phi_n(x_n)\vac
\ee
and use the translation properties (\ref{5h7}) and (\ref{5h8}),
\bea {\cal W} &=& \sum_{I_1\ldots I_{n-1}} e^{-ip_1\cdot(x_1-x_2)}
 e^{-ip_2\cdot(x_2-x_3)} \ldots e^{-ip_{n-1}\cdot(x_{n-1}-x_n)}\nonumber \\
 & &\hspace*{20mm}\times\ \ \Vac\phi_1(0)\ket{I_1}\bra{I_1}\phi_2(0)\ket{I_2}
             \ldots\bra{I_{n-1}}\phi_n(0)\vac
\label{5i6}  \eea
where $(p_k)_\mu$ is the appropriate four-momentum eigenvalue:
\be  P_\mu\ket{I_k} = (p_k)_\mu\ket{I_k}
%\label{5i7}
\ee
Each summation $\sum_{I_k}$ includes an integral over $(p_k)_\mu$ which
has to be regulated at large momenta, as in Eq.\ (\ref{4g1}). Each
coordinate $(x_k)_\mu$ is complexified
\be  x_k \longrightarrow z_k  \ee
such that successive differences
\be  z_k - z_{k+1} = x_k - x_{k+1} - i\eta_k\ ,
                    \ \  k = 1,2, \ldots , n-1
\label{5i8}  \ee
acquire imaginary parts $-\eta_k$ with each real four-vector
$(\eta_k)_\mu$ restricted to lie within the forward light cone. This
ensures that each oscillatory exponential in (\ref{5i6}) acquires a
damping factor at large $p_k$:
\be e^{-ip_k\cdot(x_k-x_{k+1})} \longrightarrow
 e^{-ip_k\cdot(x_k-x_{k+1})} e^{-p_k\cdot\eta_k}\ ,\ \ p_k\cdot\eta_k > 0
%\label {5i9}
\ee
Then the unordered function (\ref{5h9}) is obtained as a boundary value
\be {\cal W}(x_1-x_2,x_2-x_3,\ldots ,x_{n-1}-x_n)
 = \begin{array}{c}\lim \\[-1.2mm] \eta_1\ldots\eta_{n-1}
            \rightharpoonup 0 \end{array} W(z_1,z_2,\ldots ,z_n)
%\label{5j1}
\ee
of a function $W$ of several complex variables:
\bea \lefteqn{W(z_1,z_2,\ldots ,z_n)} \nonumber \\
&=& \sum_{I_1\ldots I_{n-1}}
           \Vac\phi_1(0)\ket{I_1}\bra{I_1}\phi_2(0)\ket{I_2}
     \ldots\bra{I_{n-1}}\phi_n(0)\vac e^{-i\sum_k p_k\cdot(z_k-z_{k+1})}
\label{5j2}  \eea

Note that vacuum expectation values of field products $\Pi_j\phi_j(x_j)$
written in various orders correspond to different boundary values of
the same complex function (\ref{5j2}). To obtain
\be \Vac\phi_{j_1}(x_{j_1})\phi_{j_2}(x_{j_2})\ldots\phi_{j_n}(x_{j_n})\vac
\label{5j3}  \ee
simply replace Eq.\ (\ref{5i8}) by
\be z_{j_k} - z_{j_{k+1}} = x_{j_k} - x_{j_{k+1}} - i\eta_k\ ,\ \
    \eta^2_k > 0\ ,\ \eta^0_k > 0
%\label{5j4}
\ee
and take the limit $\eta_1\ldots\eta_n\rightharpoonup 0$ of
$W(z_1\ldots z_n)$.

If neighbouring operators in Eq.\ (\ref{5j3}) are space-like separated and
so commute (or anti-commute), the two boundary values corresponding to their
interchange coincide. The absence of any discontinuity indicates that $W$
is analytic in the corresponding $z$ variables. Indeed, the space-like
region defined by
\be \left(x_j - x_k\right)^2 < 0\ , \ \ \mbox{\rm all}\ j,k
\label{5j5}  \ee
lies within the domain of analyticity ${\cal D}$ of $W$ as a function of
$z_1,z_2, \ldots ,z_n$.

The space-like region (\ref{5j5}) can be extended to a special subspace
of ${\cal D}$ called the {\ital Euclidean region\/}. It consists of
non-coincident points $\{z_1,\ldots,z_n\}$ for which each $z_k$ has
real space components and pure imaginary time components:
\be z_k \longrightarrow z^E_k = \left(-i\x_4,{\bf x}\right)_k\ ,\ \
                                 z^E_j \not= z^E_k
%\label{5j6}
\ee
The real variables ${\bf x}_k$ and $(\x_k)_4$ can be treated as the
components of a four dimensional Euclidean vector $(\x_k)_\mu ,\
\mu = 1,2,3,4$:
\bea  \x_k &=& \left({\bf x}, \x_4\right)_k \nonumber \\
  \left(\x_k\right)_\mu\cdot\left(\x_k\right)_\mu
  &=& {\bf x}^2_k + \left(\x_k\right)^2_4\ =\ - z^E_k \cdot z^E_k
%\label{5j7}
\eea
Amplitudes $W$ evaluated at these points are called Euclidean Green's
functions, or Schwinger functions:
\be {\cal G}(\x_1,\ldots,\x_n) = W\left(z^E_1,\ldots,z^E_n\right)
\label{5j8}  \ee

These functions can be regarded as amplitudes of a Euclidean version of the
theory in which Euclidean field operators $\Phi_j(\x_j)$ act on a state
vector space distinct from but analogous to that for the Minkowskian
theory. In particular, there is a Euclidean ``vacuum'' state $\Evac$
distinguished by its invariance under Euclidean translations and $SO(4)$
rotations. Euclidean Green's functions are given by vacuum expectation
values
\be {\cal G}(\x_1,\ldots,\x_n)
    = \EVac\Phi_1(\x_1)\ldots\Phi_n(\x_n)\Evac
\label{5j9}  \ee
which depend on Euclidean coordinate differences $\x_1-\x_2, \ldots ,
\x_{n-1}-\x_n$. Because of analyticity, there are no discontinuities
associated with different operator orderings, so we have
\be [\Phi_j(\x_j),\Phi_k(\x_k)]_\mp = 0
\label{5k1}  \ee
{\ital throughout\/} the Euclidean region. It is this feature which makes
Euclidean operators easier to handle than their Minkowskian counterparts.

The commutativity property (\ref{5k1}) does not mean that the Euclidean
theory is classical. Its quantum nature becomes evident when Euclidean
Green's functions ${\cal G}$ are extended to include coincident points
$\x_j=\x_k$, which lie on the boundary of the analyticity domain
${\cal D}$.

For example, consider the free two-point function (\ref{5i1}) with mass
$m=0$ continued to the Euclidean region. From Eq.\ (\ref{4g2}), we find
\be {\cal G}(\x-\y) = \EVac\Phi(\x)\Phi^\adj(\y)\Evac
                     = \frac{1}{4\pi^2(\x-\y)^2}
\label{5k2}  \ee
The equation of motion satisfied by the free Euclidean field $\Phi(\x)$
is
\be  \del^2\Phi
    \equiv \frac{\del}{\del\x_\mu}\frac{\del}{\del\x_\mu}\Phi = 0
%\label{5k3}
\ee
but for the amplitude (\ref{5k2}), we find
\be \del^2{\cal G}(\x-\y) = - \delta^4(\x-\y)
%\label{5k4}
\ee
instead. That is why the term ``Green's function'' is appropriate
for Euclidean amplitudes (\ref{5j8}).

There is a close resemblance between Euclidean Green's functions and
time-ordered vacuum expectation values in the Minkowskian theory, which
are also Green's functions. In time-ordered products, each operator
appears to the left of operators evaluated at earlier times and to the
right of those at later times. Formally, this can be indicated by
multiplying various orderings by step functions in time
differences\footnote{This works if the unordered functions are not too
singular at short distances, i.e. as various subsets of coordinates
become coincident; for example, two-point behaviour should be less
singular than $(x-y)^{-4}$. Otherwise, a renormalized time ordering
must be defined for each amplitude.}
and summing all possibilities (with a minus sign for each fermion
interchange). Thus for two operators, we have
\be T\{\phi_1(x_1)\phi_2(x_2)\}
  =  \theta(x^0_1 - x^0_2)\phi_1(x_1)\phi_2(x_2) \pm
        \theta(x^0_2 - x^0_1)\phi_2(x_2)\phi_1(x_1)
\label{5k5}  \ee
where the minus sign applies if both $\phi_1$ and $\phi_2$ are
fermionic. Then the interchangeability of Euclidean operators
$\Phi_j(\x_j)$ in (\ref{5j9}) (due to Eq.\ (\ref{5k1})) is matched by
a similar interchangeability of Minkowskian operators $\phi_j(x_j)$
in the corresponding time-ordered amplitudes
\be T\Vac\phi_1(x_1)\phi_2(x_2)\ldots\phi_n(x_n)\vac
%\label{5k6}
\ee

The time-ordered amplitude corresponding to (\ref{5k2})
\be T\bra{0}\phi(x)\phi^\adj(y)\ket{0}
  = \theta(x^0 - y^0)i\Delta^+(x-y) + \theta(y^0 - x^0)i\Delta^+(y-x)
%\label{5k7}
\ee
can be deduced from the result (\ref{4g3}) for $\Delta^+(x)$:
\bea T\bra{0}\phi(x)\phi^\adj(y)\ket{0} &=& -\frac{1}{4\pi^2}
  \left(\frac{\theta(x^0 - y^0)}{(x-y)^2-i\epsilon(x^0 - y^0)}
     +  \frac{\theta(y^0 - x^0)}{(y-x)^2-i\epsilon(y^0 - x^0)}\right)
                                  \nonumber  \\
  &=& -\frac{1}{4\pi^2\left((x-y)^2 - i\epsilon\right)}
\label{5k8}  \eea
This is the coordinate-space representation of the Feynman propagator
for a massless scalar field. Note the simple substitution
$\x^2 \leftrightarrow -x^2$ which relates Eqs.\ (\ref{5k2}) and (\ref{5k8}).

The $-i\epsilon$ prescription for the singularity in (\ref{5k8})
is responsible for it behaving as a Green's function:
\be \del^2 T\bra{0}\phi(x)\phi^\adj(y)\ket{0} = -i\delta^4(x-y)
\label{5k9}  \ee
This should be compared with the result (\ref{4g4}) for $\Delta^+$.
Note that we could have deduced (\ref{5k9}) by direct differentiation
of the definition (\ref{5k5}), taking into account the identity
$\del_\mu\theta(x^0) = g_{\mu 0}\delta(x^0)$ and the equation of motion
$\del^2\phi = 0$ and canonical commutator (\ref{5g1}) for the free scalar
field $\phi$:
\be \del^2_x T\{\phi(x)\phi^\adj(y)\}
   = [\del_0\phi(x),\phi^\adj(y)] \delta(x_0 - y_0) = -i\delta^4(x-y)
%\label{5m1}
\ee

The relation between time-ordered and Euclidean amplitudes is equally
direct in momentum space. In Minkowski space, the propagator (\ref{5k8})
can be written
\be T\bra{0}\phi(x)\phi^\adj(y)\ket{0}
  = \loopint{k}\frac{i}{k^2 + i\epsilon} e^{-ik\cdot(x-y)}
%\label{5m2}
\ee
and in the massive case, we get
\be \frac{i}{k^2 + i\epsilon}
        \longrightarrow \frac{i}{k^2 - m^2 + i\epsilon}
\label{5m3}  \ee
instead. The continuation to Euclidean space $k^2 \rightarrow
- {\rm k}^2$ permits the on-shell singularity at $k^2 = m^2$
to be avoided.

Perturbative amplitudes are essentially four-momentum integrals over
products of the Feynman propagators (\ref{5m3}), so it was
natural that the idea to continue to Euclidean space should have arisen
first in that context \cite{DWW}. The formulation of Euclidean theory
in terms of operators was first suggested by Schwinger \cite{Schwinger2},
and developed as an alternative approach to axiomatic theory by
Symanzik \cite{Symanzik2} for many years before it became fashionable.

The foregoing discussion dodges a good deal of very advanced analysis
of interest to rigorous mathematical physicists. Devotees should consult
the book by Simon \cite{Simon}.
\sect{6}{Functional Methods}

Feynman's method of ``path'' or ``functional'' integration \cite{Feynman}
is based on an ingenious use of quantum superposition combined with some
prescient remarks of Dirac \cite{Dirac3} about the role of the action
in quantum theory.

In ordinary quantum mechanics, as a system evolves in time $t$, it passes
through a variable mixture of eigenstates of a given dynamical
coordinate $Q(t)$. This corresponds to the fact that generally, the
uncertainty $\Delta Q(t)$ is not zero; at each time $t$, a range of
eigenvalues $\{q_n(t)\}$ of $Q(T)$ is involved. Therefore, the evolution
of the system can be viewed as consisting of many sequences of hops from
eigenvalue to eigenvalue
\be  q_{n_0}(t_0) \rightarrow q_{n_1}(t_1) \ldots
  \rightarrow q_{n_\ell}(t_\ell)
%\label{6a1}
\ee
over a large number $\ell$ of small positive increments of time
$t_k \rightarrow t_{k+1}$. Each sequence of hops can be considered to
define a path $q(t)$. Then the full amplitude is given by an appropriate
sum over all such paths, or path integral:
\be  \mbox{\rm amplitude} = \int\![dq(t)]\,\{\mbox{\rm integrand}\}
\label{6a2}  \ee
It turns out that the integrand is just $\exp i\{\mbox{\rm Action}/\hbar\}$.
A detailed account can be found in the book by Feynman and Hibbs
\cite{Feynman}.

In field theory, the path integral becomes a sum over functions $\varphi$
of $x_\mu$ or its Euclidean cousin $\x_\mu$. We will consider the Euclidean
case, where (for bosons at least) the commutativity property
(\ref{5k1}) allows {\ital simultaneous diagonalization\/} of the set
of Euclidean field operators $\Phi_j(\x)$ labelled by {\ital all\/}
$\x_\mu$. This yields eigenvalues $\varphi_j(\x)$ which are also
labelled by $\x_\mu$:
\be \Phi_j(\x)\ket{\varphi_1, \ldots ,\varphi_n}
                 = \varphi_j(\x)\ket{\varphi_1, \ldots ,\varphi_n}
\label{6a3}  \ee
It is these eigenvalues over which the functional summation is to be
performed.

As in ordinary calculus, derivatives are easier than integrals,
so we tackle functional differentiation first, cover path integration
for bosons, and finally, deal with fermions, where a special treatment
is necessary \cite{Berezin}.

Functional integrals have played an important role in the development
of gauge theories, and so they feature prominently in modern textbooks
[9, 49--53].\\[4mm]
{\ital 6.1 Functional \vspace{1mm}Differentiation}

A functional $F$ maps functions $\varphi(x)$ to numbers $F[\varphi]$:
\be  \varphi(x) \stackrel{F}{\longrightarrow} F[\varphi]
%\label{6a4}
\ee
For example, for real functions $\varphi$, we can construct
functionals such as
\be  F[\varphi] = \Int{4}{x}\,\varphi^2(x)
\label{6a5}  \ee
or
\be  F[\varphi] = \Int{4}{x}\,\left(\del\varphi(x)\right)^2
\label{6a6}  \ee
which are quadratic in $\varphi$. Generalized functions correspond
to functionals {\ital linear\/} in the test function; the smeared operator
(\ref{4p}) is an example.

The functional derivative of $F$ is defined as the generalized function
$\delta F/\delta\varphi(x)$ smeared over test functions $f(x)$
\be \Int{4}{x}\, f(x)\frac{\delta F[\varphi]}{\delta\varphi(x)}\,
 = \begin{array}{c}\lim \\[-1.2mm] \epsilon \rightarrow 0 \end{array}
   \frac{F[\varphi + \epsilon f] - F[\varphi]}{\epsilon}
%\label{6a7}
\ee
whenever the limit exists. Thus, for the example (\ref{6a5}), we can
calculate
\bea F[\varphi + \epsilon f]
 &=& \Int{4}{x}\,\{\varphi^2 + 2\epsilon f\varphi + \epsilon^2 f^2\}
                                        \nonumber  \\
 &=& F[\varphi] + 2\epsilon\!\Int{4}{x}\, f(x)\varphi(x) + O(\epsilon^2)
%\label{6a8}
\eea
and so conclude
\be \frac{\delta F[\varphi]}{\delta\varphi(x)} = 2\varphi(x)
%\label{6a9}
\ee
Similarly, for the example (\ref{6a6}), we have
\be F[\varphi + \epsilon f] - F[\varphi]
 = 2\epsilon\!\Int{4}{x}\, (\del f)\cdot(\del\varphi) + O(\epsilon^2)
 = -2\epsilon\!\Int{4}{x}\, f \del^2 \varphi + O(\epsilon^2)
%\label{6b1}
\ee
and hence
\be \frac{\delta F[\varphi]}{\delta\varphi(x)} = -2\del^2 \varphi(x)
%\label{6b2}
\ee

Sometimes it is stated that the functional derivative can be defined as
follows:
\be \frac{\delta F[\varphi(y)]}{\delta\varphi(x)}\, \stackrel{?}{=}
  \begin{array}{c}\lim \\[-1.2mm] \epsilon \rightarrow 0 \end{array}
  \frac{F[\varphi(y) + \epsilon\delta^4(x-y)] - F[\varphi(y)]}{\epsilon}
%\label{6b3}
\ee
The trouble with this idea can be seen from the example (\ref{6a5}), for
which we would have the ill-defined expression
\be F[\varphi(y) + \epsilon\delta^4(x-y)] \stackrel{?}{=}
      \Int{4}{y}\,\left(\varphi(y) + \epsilon\delta^4(x-y)\right)^2
%\label{6b4}
\ee

As in ordinary calculus, functional differentiation can be automated
by combining basic results such as
\bea
\frac{\delta\varphi(y)}{\delta\varphi(x)} &=& \delta^4(x-y)  \nonumber \\
\frac{\delta}{\delta\varphi(x)}
         \left\{\frac{\del}{\del y_\mu}\varphi(y)\right\}
 &=& \frac{\del}{\del y_\mu}\delta^4(x-y)
%\label{6b5}
\eea
with the product rule
\be \frac{\delta (FG)}{\delta\varphi}
 = \frac{\delta F}{\delta\varphi}G + F\frac{\delta G}{\delta\varphi}
%\label{6b6}
\ee
and the chain rule
\be \frac{\delta}{\delta\psi(x)}  =  \Int{4}{y}\,
  \frac{\delta\varphi(y)}{\delta\psi(x)}\frac{\delta}{\delta\varphi(y)}
%\label{6b7}
\ee
Thus functional derivatives are as easy to calculate as ordinary
derivatives.

Another rule worth noting is the formula for translations in function space:
\be F[\varphi + \xi]
 = \exp\left\{\Int{4}{x}\,\xi(x)\frac{\delta}{\delta\varphi(x)}
       \right\} F[\varphi]
%\label{6b8}
\ee
This is a functional version of the Taylor expansion for ordinary
functions $f$:
\be f(x+a) = e^{a\cdot\del}f(x)
%\label{6b9}
\ee

Functional derivatives occur naturally in field theories. The
standard problem is that of determining the classical equation of
motion by varying the action $S$ and setting the result equal to zero.
This can be done either directly, by treating $S$ as a functional of
$\varphi$, or via the Euler-Lagrange formalism, where the corresponding
Lagrangian ${\cal L}$ is treated as a function of $\varphi$ and
$\del\varphi$, in the {\ital ordinary\/} sense of the term ``function'':
\be S[\varphi] = \Int{4}{y}\,{\cal L}\big(\varphi(y),\del\varphi(y)\big)
%\label{6c1}
\ee
Note the use of square and curved brackets to distinguish dependence
as a functional from that as a function. A relation between the two
types of derivative is readily deduced:
\be \frac{\delta S[\varphi]}{\delta\varphi(x)}
 = \frac{\del{\cal L}}{\del\varphi}
      - \del_\mu\frac{\del{\cal L}}{\del\del_\mu\varphi}
\label{6c2}  \ee
At a minimum of $S$, we get the
classical equation of motion either in Euler-Lagrange form (the right-hand
side of (\ref{6c2}) vanishing), or more simply, in functional form:
\be \frac{\delta S[\varphi]}{\delta\varphi(x)} = 0
\label{6c3}
\ee

A formalism for the systematic use of functional derivatives in
quantum field theory was developed by Schwinger \cite{Schwinger3}.\\[4mm]
{\ital 6.2 Functional Integrals and the \vspace{1mm}Action}

Consider a single Euclidean Bose field $\Phi$. Because of the commutativity
property (\ref{5k1}), we argue as in Eq.\ (\ref{6a3}) that $\Phi =
\Phi(\x)$ can be treated as a collection of mutually commuting operators
labelled by all Euclidean points $\x_\mu$. Therefore, we should be able
to construct simultaneous eigenstates $\ket{\varphi}$ of this
operator set:
\be \Phi(\x)\ket{\varphi} = \varphi(\x)\ket{\varphi}
\label{6c4}
\ee
The symbol $\varphi$ inside $\ket{\varphi}$ represents the set of
eigenvalues $\{\varphi(\x),\ \mbox{\rm for all}\ \x_\mu\}$.
Presumably, these eigenstates span state-vector space, so there should
be a completeness relation of the form
\be \int\![d\varphi]\, \ket{\varphi}\bra{\varphi} = 1
\label{6c5}
\ee
where $\int\![d\varphi]$ indicates that a sum over functions $\varphi$,
or functional integral, is to be performed.

This statement of completeness specifies that we are dealing
with a field theory. A less dense state structure, such as that corresponding
to eigenstates of the quantum coordinate $Q(t)$, would correspond to
quantum mechanics. If a richer state structure is desired, eigenstates
of string operators should be considered.

Since eigenstates are a functional $\ket{\varphi}$ of $\varphi$, their
orthonormality must be expressed in terms of a delta functional:
\be \braket{\varphi_1}{\varphi_2} = \delta[\varphi_1 - \varphi_2]
%\label{6c6}
\ee
The delta functional should have the property
\be \int\![d\varphi_1]\,\delta[\varphi_1 - \varphi_2] f[\varphi_1]
               =  f[\varphi_2]
%\label{6c7}
\ee
As a check, note that this is consistent with choosing a functional $F$
\[ F[\varphi] = \braket{\varphi}{f} \]
and applying completeness:
\[ \int\![d\varphi_1]\, \braket{\varphi_2}{\varphi_1}
              \braket{\varphi_1}{F} = \braket{\varphi_2}{F}  \]

Now consider the insertion of completeness relations around each operator
in an $n$-point Euclidean Green's function:
\bea \lefteqn{\EVac\Phi(\x_1)\ldots\Phi(\x_n)\Evac} \nonumber  \\
& &=\ \int\![d\varphi_1]\ldots\int\![d\varphi_{n+1}]
      \braket{\mbox{\small\sc vac}}{\varphi_1}
      \bra{\varphi_1}\Phi(\x_1)\ket{\varphi_2}\ldots
      \bra{\varphi_n}\Phi(\x_n)\ket{\varphi_{n+1}}
      \braket{\varphi_{n+1}}{\mbox{\small\sc vac}}
%\label{6c8}
\eea
Eqs.\ (\ref{6c4}) and (\ref{6c5}) imply
\be \bra{\varphi_1}\Phi(\x_1)\ket{\varphi_2}
  = \varphi_2(\x_1)\braket{\varphi_1}{\varphi_2}
  = \varphi_2(\x_1)\delta[\varphi_1 - \varphi_2]
%\label{6c9}
\ee
so all except one of the $[d\varphi_k]$ integrals is trivial:
\be \EVac\prod_{j=1}^{n}\{\Phi(\x_j)\}\Evac
 =  \int\![d\varphi]\,|\braket{\mbox{\small\sc vac}}{\varphi}|^2
         \prod_{j=1}^{n} \varphi(\x_j)
%\label{6d1}
\ee

The probability $|\braket{\mbox{\small\sc vac}}{\varphi}|^2$ is a
positive functional of $\varphi$. It is convenient to write it in
the following form,
\be |\braket{\mbox{\small\sc vac}}{\varphi}|^2
   = {\cal N}\,\exp -S[\varphi]/\hbar
%\label{6d2}
\ee
where ${\cal N}$ is a normalization factor chosen to ensure unit
normalization for the Euclidean vacuum:
\be \braket{\mbox{\small\sc vac}}{\mbox{\small\sc vac}} = 1
%\label{6d3}
\ee
The constant $\hbar$ is the usual quantum of action, so $S[\varphi]$ has
the dimensions of action; indeed, it will be shown to be a Euclidean
version of the action. Note that we have a real positive weighting
factor $\exp -S/\hbar$ instead of the oscillatory factor $\exp iS/\hbar$
obtained in Minkowski space.

For the moment, we treat $S[\varphi]$ as a functional
which characterises a given field theory. With these definitions,
we find:
\be \EVac\prod_{j=1}^{n}\{\Phi(\x_j)\}\Evac
 =  \frac{\int\![d\varphi]\,\prod_{j=1}^{n}\{\varphi(\x_j)\}
            \exp -S[\varphi]/\hbar}
         {\int\![d\varphi]\,\exp -S[\varphi]/\hbar}
\label{6d4}  \ee
Notice that the result is a {\ital functional average\/} of\ \,
$\prod_{j=1}^{n}\varphi(\x_j)$.

The discussion above makes no reference to the origin in function space
$\{\varphi\}$. Our choice of operator $\Phi(\x)$ was arbitrary; instead,
we could have chosen to consider the eigenvalue problem for
\[  \widetilde{\Phi}(\x) = \Phi(\x) + f(x)I  \]
for any function $f(x)$. Therefore functional integrals should have
a translation property
\be \int\![d\varphi]\,F[\varphi + f] = \int\![d\varphi]\,F[\varphi]
\label{6d5}  \ee
analogous to that for ordinary infinite integrals:
\[ \int^\infty_{-\infty}\! dx\, f(x+a)
             = \int^\infty_{-\infty}\! dx\, f(x)   \]
When Eq.\ (\ref{6d5}) is expanded in $f$, the linear term is
\be \int\![d\varphi]\,\frac{\delta}{\delta\varphi} F[\varphi] = 0
\label{6d6}  \ee

We are now in a position to identify $S[\varphi]$ as an action. In Eq.\
(\ref{6d6}), let the functional $F$ be given by the integrand of the
numerator of Eq.\ (\ref{6d4}):
\be F[\varphi]
   = \prod_{j=1}^{n}\{\varphi(\x_j)\} \exp -S[\varphi]/\hbar
%\label{6d7}
\ee
Its functional derivative is
\be \frac{\delta F\rule{1mm}{0mm}}{\delta\varphi(\x)}
 = \left\{- \frac{1}{\hbar} \frac{\delta S\rule{1mm}{0mm}}{\delta\varphi(\x)}
            \prod_{j=1}^{n}\{\varphi(\x_j)\}
 + \sum_k\delta^4(\x-\x_k)\prod_{j\not= k}\varphi(\x_k)\right\}
                                      \exp -S[\varphi]/\hbar
%\label{6d8}
\ee
If we integrate over $\varphi$ and use Eq.\ (\ref{6d6}), the result is
a set of Schwinger-Dyson equations:
\be
\EVac\,\frac{\delta S[\Phi]}{\delta\Phi(\x)}\prod_{j=1}^{n}\{\Phi(\x_j)\}\Evac
= \hbar\sum_k\delta^4(\x-\x_k)\EVac\prod_{j\not= k}\{\Phi(\x_k)\}\Evac
\label{6d9}  \ee
In the classical limit $\hbar\rightarrow 0$, we find that the variation
of $S$ vanishes exactly, as in Eq.\ (\ref{6c3}); the contact terms on
the right-hand side of Eq.\ (\ref{6d9}) are evidently quantum corrections.
We conclude that $S[\varphi]$ is the classical action functional, apart
from possible $O(\hbar)$ corrections.

It is convenient to summarize results obtained for the set of all Green's
functions (\ref{6d4}) by introducing the generating functional
\be Z[j] = \int\![d\varphi]\,\exp -\Big\{S[\varphi]
                      + \Int{4}{\x}\, j(\x)\varphi(\x)\Big\}
\label{6e1}  \ee
Variations with respect to the ``source'' function $j(x)$ result
in insertions of the operator $\Phi(x)$:
\be \EVac\prod_{j=1}^{n}\{\Phi(\x_j)\}\Evac  = \left.\frac{1}{Z[0]}
    \prod_k\left\{-\frac{\delta\rule{2mm}{0mm} }{\delta j(\x_k)}\right\}
                                 Z[j]\right|_{j=0}
%\label{6e2}
\ee

For future reference, we note the identity
\be F\left[-\frac{\delta}{\delta j}\right]
                     \exp\, -\!\Int{4}{\x}\, j(\x) \varphi(\x)
 = F[\varphi]\,\exp\, -\!\Int{4}{\x}\, j(\x) \varphi(\x)  \vspace{4mm}
\label{6e3}  \ee
which can be verified by expanding the functional $F$ as a Taylor
series.\\[4mm]
{\ital 6.3 Evaluating Functional \vspace{1mm}Integrals}

So far, we have not tried to interpret the meaning of the ``integral
over $\varphi$''.

In the lattice approach \cite{Wilson2}, the coordinates $\x_\mu$
are restricted to discrete sites $(\x_\mu)_n$ so that integrals can be
replaced by sums over discrete variables $\varphi_n$:
\be \int\![d\varphi] = \prod_n\int^\infty_{-\infty}\! d\varphi_n\ ,\ \
\varphi_n = \varphi(\x_n)
%\label{6e4}
\ee
This approximation introduces a lattice spacing $a$. Derivatives
$\del\varphi$ are approximated by differences $(\varphi_{n+1} -
\varphi_n)/a$. It is assumed that the correct result is obtained
in the limit $a\rightarrow 0$.

As first noted by Symanzik \cite{Symanzik2}, the discrete version of
a Euclidean path integral has the structure of a partition function
in statistical mechanics:
\be  Z = \mbox{\rm Tr}\, \exp -\beta H
%\label{6e5}
\ee
Consequently, techniques from statistical mechanics can be applied
to problems in quantum field theory \cite{JZJ,Wilson2}.

The main alternative to the lattice approach is ``Gaussian'' integration.
This leads directly to the rules for Feynman diagrams. As a simple
example, we consider a Hermitian Bose field operator $\Phi(\x)$ with real
eigenvalues $\varphi(\x)$.

Let us write each function $\varphi(\x)$ as a linear combination
\be \varphi(\x) = \sum_n z_n\,\varphi_n(\x)\
%label{6e6}
\ee
of complete orthonormal functions $\varphi_n(\x)$ with real coefficients
$z_n$:
\bea \Int{4}{\x}\,\varphi^\star_m(\x)\varphi_n(x) &=& \delta_{mn}
                                               \nonumber \\
    \sum_n \varphi_n(\x)\varphi^\star_n(\y) &=& \delta^4(\x-\y)
%\label{6e7}
\eea
Note that $\varphi_n$ may be complex, as in the Fourier analysis of a
real function.

The set
\be \{\varphi_n(\x)\, ,\ n=1,2,3,\ldots\}
\label{6e8}  \ee
can be regarded as defining orthogonal vectors with unit length in
function space. As the parameters $z_n$ are varied over all real values,
all functions $\varphi$ are produced. Thus
\[ \{z_1, z_2, z_3, \ldots \}  \]
is a set of coordinates for an infinite-dimensional Cartesian space,
where each point labels a function $\varphi$. The measure for path
summation is given by a volume element in $\{z_n\}$ space:
\be [d\varphi] = \prod_n\left(dz_n/\sqrt{2\pi}\right)
%\label(6e9}
\ee
This prescription does not depend on the choice of complete set
(\ref{6e8}), provided that boundary conditions defining the function
space $\{\varphi\}$ are respected. The normalization $1/\sqrt{2\pi}$
for each $z$ integral is a convention; such factors cancel in the
expression (\ref{6d4}) for Euclidean Green's functions.

The standard Gaussian integral is
\be {\cal I} = \int\![d\varphi]\, \exp -\half\Int{4}{\x}\,
                     \varphi(\x)\D \varphi(\x)
\label{6f1}  \ee
where $\D$ is a differential operator. It is convenient to choose
the eigenfunctions of $\D$ as a basis (\ref{6e8}) for function space:
\be \D \varphi_n(\x) = E_n\varphi_n(\x)
%\label{6f2}
\ee
Then Eq.\ (\ref{6f1}) becomes a product of ordinary Gaussian integrals,
each of the form
\[  \int^\infty_{-\infty}\! dz\,\exp -\half Ez^2 = \sqrt{2\pi/E}  \]
So we find
\be {\cal I} = \prod_n\left\{\int^\infty_{-\infty}\! \left(dz_n/
                  \sqrt{2\pi}\right)\exp -\half E_n z^2_n\right\}
             = (\det D)^{-1/2}
\label{6f3}  \ee
where the determinant of the operator $\D$ is defined to be the product
of its eigenvalues:
\be  \det D = \prod_n E_n
\label{6f4}  \ee

This example can be easily generalized to include a source $j(\x)$ for
$\varphi$:
\be {\cal I}[j] = \int\![d\varphi]\, \exp -\Int{4}{\x}\,
          \{\half\varphi(\x) \D \varphi(\x) + j(\x)\varphi(\x)\}
%\label{6f5}
\ee
A shift of integration variable
\be  \varphi \longrightarrow \varphi + \D^{-1} j
%\label{6f6}
\ee
yields the result
\bea {\cal I} &=& (\det D)^{-1/2} \exp\half\Int{4}{\x}\,
                                j(\x) \D^{-1} j(\x)  \nonumber \\
 &=& (\det D)^{-1/2} \exp\half\Int{4}{\x}\Int{4}{\y}\,
                                j(\x) G(\x,\y) j(\y)
%\label{6f7}
\eea
where
\be  G(\x,\y) = \sum_n E^{-1}_n \varphi_n(\x)\varphi^\star_n(\y)
              = \D^{-1} \delta(\x-\y)
\label{6f8}  \ee
is the propagator of the operator $\D$.

The identity (\ref{6e3}) permits an immediate generalization of eq.\
(\ref{6f8}) to include a potential $V[\varphi]$:
\be \int\![d\varphi]\, \exp -\Int{4}{\x}\,
          \{\half\varphi(\x) \D \varphi(\x) + j\varphi + V[\varphi]\}
 = \exp - \Int{4}{\x}\, V[-\delta/\delta j(x)]\,{\cal I}[j]
\label{6f9}  \ee

These expressions can be used to generate Feynman rules very efficiently;
(indeed, that is essentially how Feynman arrived at them originally).
For example, consider the Euclidean version of the theory (\ref{5e3}):
\be S[\varphi]
= \Int{4}{\x}\,\{\half\varphi(-\del^2 + m^2)\varphi + \lambda\varphi^4\}
%\label{6h1}
\ee
Then, according to Eq.\ (\ref{6f9}), the generating functional $Z[j]$ of
(\ref{6e1}) satifies the formula
\be Z[j]/Z[0] =
  \exp\left\{-\lambda\Int{4}{\x}\,\big(\delta/\delta j(\x)\big)^4\right\}\,
  \exp\half\Int{4}{\x}\Int{4}{\y}\, j(\x) G(\x-\y) j(\y)
\label{6h2}  \ee
where
\be G(\x-\y) = (-\del^2 + m^2)^{-1}_{\mbox{\eightrm x}} \delta^4(\x-\y)
   = \int\!\frac{d^4{\rm k}}{(2\pi)^4}
            \frac{\exp -i{\rm k}\cdot(\x-\y)}{{\rm k}^2 + m^2}
%\label{6h3}
\ee
is the Euclidean Feynman propagator. In Feynman diagrams, $G(\x-\y)$ is
represented by a line joining $\x$ and $\y$. The second exponential
factor in (\ref{6h2}) provides any number of propagator lines, which are
then joined at their ends by the action of the first exponent. Each
time the exponent $-\lambda\Int{4}{\x}\,\big(\delta/\delta j(\x)\big)^4$
acts, a four-legged vertex is formed. Feynman diagrams are staple fare
in textbooks, so there is little point trying to cover them further in
these brief notes.

There are two main problems with the formalism developed above. First,
there may be zero modes $E_k = 0$ which prevent $\D$ from being
inverted, especially if the zero eigenvalue is discrete. In that
case, the corresponding integrals over $z_k$ must be isolated and
handled separately via the method of ``collective coordinates''
\cite{collect}. For continuous eigenvalues, it depends on questions
of measure. For example, the operator $\D = -\del^2$ has eigenfunctions
\be u_\q(\x) = \exp -i\q\cdot\x
%\label{6g0}
\ee
labelled by the continuous four-vector $\q_\mu$, with eigenvalue $\q^2$.
There are degenerate zero modes $\q^2 = 0$, but the propagator
\be \D^{-1}\delta^4(\x-\y)
  = \int\!\frac{d^4\q}{(2\pi)^4}\frac{1}{\q^2}\exp\, -i\q\cdot(\x-\y)
  = \frac{1}{4\pi^2(\x-\y)^2}
%\label{6g1}
\ee
is well defined. On the other hand, gauge theories have an enormous
symmetry which produces a functionally degenerate set of zero modes. A
propagator can be constructed only if contributions from zero
modes are restricted by fixing the gauge sufficiently.

The other problem concerns ultra-violet infinities, which have to be
renormalized. For example, products $\prod_n$ and sums $\sum_n$ may
blow up for large eigenvalues $E_n,\ n\rightarrow\infty$. The product
(\ref{6f4}) defining $\det D$ is invariably divergent. One way of
controlling it is to use the ``zeta-function'' method \cite{Hawking}.
It makes the replacement
\be \prod_n E_n = \exp\sum_n \ln E_n \longrightarrow
    \exp -\frac{\del}{\del s}\sum_n E^{-s}_n
%\label{6g2}
\ee
where $s$ is a complex regulator. The resulting sum converges for
$\mbox{\rm Re}\,s$ sufficiently large,
\be \sum_n E^{-s}_n = \mbox{\rm Tr} D^{-s} = \zeta_D(s)
%\label{6g3}
\ee
so $\zeta_D(s)$ can be continued analytically to $s=0$:
\be \left(\det D\right)_{\mbox{\ninerm renorm}}
              = \exp -\zeta^\prime_D(0)
%\label{6g4}
\ee

Also, the operation $V[-\delta/\delta j(x)]$ in (\ref{6f9}) can
generate ultra-violet infinities at short distances by causing
too many variations $\delta/\delta j(\x)$ to act at the same point.
Loop diagrams in perturbation theory require renormalization because
of this problem.\\[4mm]
{\ital 6.4 Fermionic \vspace{1mm}Integration}

According to Eq.\ (\ref{5k1}), Euclidean fermion operators {\ital
anti\/}-commute throughout Euclidean space. Since they do not
commute, the method of simultaneous diagonalization used for bosons
in Eq.\ (\ref{6c4}) is not immediately applicable.

This problem is circumvented by a trick \cite{Berezin}.
We regard the eigenvalue problem as being formulated for operators
$\Psi(\x)$, $\bar{\Psi}(\x)$ and eigenvalues $\psi(\x)$,
$\bar{\psi}(\x)$ which take {\ital Grassmann\/} values. (Grassmann
numbers $a_1\ldots a_\ell$ are nilpotent and anticommute with each other:
$a_m^2 = 0$, $a_m a_n = a_n a_m$.)\ If the eigenvalues anticommute with
each other, simultaneous diagonalization is permitted,
\be \Psi(\x)\ket{\psi,\bar{\psi}}
         =  \psi(\x)\ket{\psi,\bar{\psi}}\ ,\ \
\bar{\Psi}(\x)\ket{\psi,\bar{\psi}}
         =  \bar{\psi}(\x)\ket{\psi,\bar{\psi}}
%\label{7a1}
\ee
where the states $\ket{\psi,\bar{\psi}}$ are not Grassmann valued,
and are assumed to span the Euclidean state vector space:
\be \int\! [d\bar{\psi}d\psi]\,
  \ket{\psi,\bar{\psi}}\bra{\psi,\bar{\psi}} = I
\label{7a2}
\ee

A Gaussian interpretation of the sum over fermionic paths in Eq.\
(\ref{7a2}) (sometimes called a ``Graussian'' integral) is possible
for integrals of Grassmann-valued functionals of $\psi,\bar{\psi}$.

First, we expand the eigenvalue functions in terms of
a complete orthonormal set of ordinary spinor functions $u_n(\x)$:
\be \psi(\x) = \sum_n \xi_n u_n(\x)\ ,\ \
    \bar{\psi}(\x) = \sum_n \eta_n u^\adj_n(\x)
%\label{7a3}
\ee
The Grassmann character of $\psi(\x)$ and $\bar{\psi}(\x)$ is
carried by the coefficients $\{\xi_n, \eta_n\}$, which are
independent Grassmann variables.

We have to interpret an integral of the form
\be  [d\bar{\psi} d\psi] = \prod_n (d\eta_n d\xi_n)
%\label{7a4}
\ee
where the differentials $d\eta_n$ and $d\xi_n$ are also Grassmannian.
This integral should be translation invariant, as in the bosonic case
(\ref{6d5}), so each Grassmann integral must also have this property,
\be \int\! d\xi\, f(\xi + \xi_0) = \int\! d\xi f(\xi)
%\label{7a5}
\ee
where $\xi_0$ is an independent Grassmann constant. Since $\xi$ is
nilpotent, the Taylor series for $f(\xi)$ terminates,
\be f(\xi) = f(0) + \xi f^\prime(0)
%\label{7a6}
\ee
so it is sufficient to determine the integrals of $\xi$ and of a
constant. So, choosing our function to be $f(\xi) = \xi$,
\be \int\! d\xi\, \xi = \int\! d\xi\, \xi + \int\! d\xi\, \xi_0
%\label{7a7}
\ee
we conclude that the only consistent answer for the integral over a
constant is zero:
\be \int\! d\xi = 0
%\label{7a8}
\ee
The only remaining integral, that over $\xi$, must give an ordinary
non-zero number which, by convention, is chosen to be unity
\be \int\! d\xi\, \xi = 1
%\label{7a9}
\ee
so that $\int\! d\xi\, f(\xi)$ is just the derivative $f^\prime(0)$ at
the origin. Whatever normalization is chosen here, it will cancel out
in the ratio (\ref{6d4}).

The fermionic Gaussian integral analogous to (\ref{6f1}) can now
be evaluated directly:
\bea {\cal I} &=& \int\! [d\bar{\psi}d\psi]\,
   \exp\, -\!\Int{4}{\x}\, \bar{\psi}(\x)\D \psi(\x) \nonumber \\
&=& \prod_n\left\{\int\!\int\! d\eta_n d\xi_n \exp -E_n\eta_n\xi_n\right\}
                                                         \nonumber \\
&=& \prod_n\left\{\int\!\int\! d\eta_n d\xi_n (1 - E_n\eta_n\xi_n)\right\}
                                                         \nonumber \\
&=& \prod_n E_n \ =\ \det D
%\label{7b1}
\eea
Note that the answer is $\det D$ instead of $(\det D)^{-1/2}$
for real boson fields (Eq.\ (\ref{6f3})), or
\be (\det D)^{-1} = \int\! [d\varphi][d\varphi^\adj]
                \exp\, -\!\Int{4}{\x}\,\varphi^\adj(\x)\D \varphi(\x)
%\label{7b2}
\ee
for complex boson fields. For Majorana fermions (mentioned in sec. 5.2),
the result is $(\det D)^{1/2}$.

A generating functional analogous to (\ref{6e1}) can be constructed if
Grassmann sources $\chi$, $\bar{\chi}$ are introduced for the
fermion fields $\bar{\psi}$, $\psi$:
\bea Z[\chi, \bar{\chi}]
&=& \int\! [d\bar{\psi} d\psi]\, \exp\, -\!\Int{4}{\x}
  \{\bar{\psi}(\x)\D\psi(\x) + \bar{\chi}\psi + \bar{\psi}\chi\}
                                      \nonumber \\
&=& \det D\; \exp \Int{4}{\x}\Int{4}{\y}\,
                 \bar{\chi}(\x) G(\x,\y) \chi(\x)
%\label{7b3}
\eea
Here the Green's function $G(\x,\y)$ is given by
\be  G(\x,\y) = \sum_n E^{-1}_n u_n(\x)u^\adj_n(\y)
              = \D^{-1} \delta(\x-\y)
\label{7b4}  \ee
instead of Eq.\ (\ref{6f8}). The Grassmann character of $\chi$ and
$\bar{\chi}$ is essential; otherwise, the antisymmetry of fermionic
Euclidean operators inside Green's functions (\ref{5j9}), as
required by Eq.\ (\ref{5k1}), would not be reproduced.

The appropriate analogue of (\ref{6f9}) allows interactions to be
introduced and hence Feynman rules to be deduced.

It may seem from this discussion that fermionic integration is not more
problematic than the bosonic version. Certainly, perturbation theory
with fermions is as straightforward as that for bosons. However the
principles behind fermionic integration are somewhat obscure. The
connection between completeness and the sum over functions is clear
for bosons, but how are we to understand Grassmann character in
the fermionic completeness relation (\ref{7a2})?

These questions may be of practical importance in non-perturbative
calculations. Certainly, it is recognised that it is difficult to
put fermions on the lattice \cite{Wilson2}. There is clearly a need
for a better understanding of fermionic integration.
\sect{7}{Outlook}

At this stage, I could have continued with a lightning introduction to
gauge theories, Fadde'ev-Popov ghosts, and Becchi-Rouet-Stora
invariance, as in the verbal version of these notes. In the event, it
became evident that a rushed version of these topics would not be of
great use here.

Indeed, I fear that these lectures have sacrificed pedagogy for brevity.
What was presented was just a rough outline, a framework on which
a satisfactory understanding of quantum field theory might be developed.
For example, I have neglected calculational techniques completely.
My excuse is that such skills take many lectures to teach, and are
covered at great length in any textbook of note.

A glance at the Table of Contents of such textbooks should provide
readers with a list of topics to pursue. A typical list could include
the following: systematic use of Feynman rules for any local
Lagrangian, calculation of low-order amplitudes in perturbation
theory, calculation of rates and cross-sections, regularization,
renormalization and the renormalization group, lattice approximation
and connection with statistical mechanics, bound-state problems,
exact and approximate symmetries, anomalies, gauge theories, quantum
chromodynamics and the standard model, supersymmetry, field-theoretic
model building, grand unification, and inevitably, strings.
\vglue 0.2cm

\end{document}